\newif\iflong
\newif\ifshort
\newif\ifexp
\newif\ifhighlight
\newif\ifnewfig
\newif\ifoldfig
\longtrue
\highlighttrue
\newfigtrue
\ifnewfig
\oldfigfalse
\else
\oldfigtrue
\fi

\iflong 
\else
\shorttrue
\fi
\documentclass[letterpaper]{article} %
\usepackage{aaai2026}  %
\usepackage{times}  %
\usepackage{helvet}  %
\usepackage{courier}  %
\usepackage[hyphens]{url}  %
\usepackage{graphicx} %
\usepackage{natbib}  %
\usepackage{caption} %
\nocopyright %

\title{How Hard is it to Explain Preferences Using Few Boolean Attributes?}
\author {
    Clemens Anzinger, 
    Jiehua Chen\textsuperscript{\rm 1},
    Christian Hatschka\textsuperscript{\rm 1},
	Manuel Sorge\textsuperscript{\rm 1},
	Alexander Temper
}
\affiliations {Insitute of Logic and Computation, TU Wien, Vienna, Austria\\
    \textsuperscript{\rm 1} \{jchen, chatschka, msorge\}@ac.tuwien.ac.at
}
\newcommand{\appendixtitle}{Supplementary Material for the Paper ``How Hard is it to Explain Preferences Using Few Boolean Attributes?''}
\usepackage{etoolbox} %
\usepackage[title]{appendix}
\newcommand{\appsymb}{$\star$}

\newcommand{\toappendix}[1]{%
  \gappto{\appendixtext}{
    {#1}
   }
}

\newcommand{\toappendixproofcontinued}[4]{%
  #1
   \gappto{\appendixtext}{
    \subsection{Continuation of the proof of \cref{#2}}\label{proof:#2}%
    \noindent{\normalfont\emph{#3}}

    {#4}
    }
}

\newcommand{\appendixproofwithstatement}[3]{%
  \gappto{\appendixtext}{
    \subsection{Proof of \cref{#1}}\label{proof:#1}
    #2 
    
    #3
  }
}

\newcommand{\appendixcorrectnessproofwithstatement}[4]{%
  #1  
  \gappto{\appendixtext}{
    \subsection{Correctness of the Construction in the Proof
      of \cref{#2}}\label{proof:#2}
    {\normalfont\emph{#3}}

    #4
    }
}

\newcommand{\appendixsection}[1]{%
  \gappto{\appendixtext}{
    \section{Additional Material for Section~\ref{#1}}
    \label{appsec:#1}
  }
}

\usepackage{amsfonts,nicefrac,amsmath,amssymb,amsthm,graphicx,booktabs}
\usepackage{mathtools,dsfont}
\usepackage{thm-restate}
\usepackage{latexsym}
\usepackage{paralist}
\usepackage{color}
\usepackage{needspace}
\usepackage[ruled,linesnumbered]{algorithm2e}

\newtheorem{theorem}{Theorem}

\newtheorem{corollary}{Corollary}

\newtheorem{observation}{Observation}
\newtheorem{claim}{Claim}[theorem]
\newtheorem{example}{Example}
\newtheorem{remark}{Remark}

\usepackage{cleveref}
\usepackage[textsize=tiny,colorinlistoftodos,textwidth=1.5cm,linecolor=green!70!black, backgroundcolor=green!10, bordercolor=black,disable]{todonotes}
\usepackage{tikz}
\usetikzlibrary{patterns}
\crefname{table}{Table}{Tables}
\crefname{figure}{Figure}{Figures}
\crefname{theorem}{Theorem}{Theorems}
\crefname{definition}{Definition}{Definitions}
\crefname{corollary}{Corollary}{Corollaries}
\crefname{observation}{Observation}{Observations}
\crefname{lemma}{Lemma}{Lemmas}
\crefname{example}{Example}{Examples}
\crefname{reduction}{Reduction}{Reductions}
\crefname{construction}{Construction}{Constructions}
\crefname{subsection}{Subsection}{Subsections}
\crefname{section}{Section}{Sections}
\crefname{proposition}{Proposition}{Propositions}
\crefname{algorithm}{Algorithm}{Algorithms}
\crefname{drule}{Rule}{Rules}
\crefname{claim}{Claim}{Claims}
\crefname{appendix}{Appendix}{Appendix}
\crefname{remark}{Remark}{Remark}

\newenvironment{claimproof}[1]
  {\begin{proof}
   \renewcommand{\qedsymbol}{\hfill~(end of the proof of~\cref{#1}~$\diamond$)}}
  {\end{proof}}

\DeclareMathOperator*{\argmax}{argmax}
\newcommand{\prob}[6]{%
  \needspace{3\baselineskip}
    \begin{description} %
      \setlength\topsep{-.15ex} \setlength\itemsep{-.2ex}
    \item[#1]
    \item[#2]#3
    \item[#4]#5
    \end{description}%
}

\newcommand{\probname}[1]{{\normalfont\textsc{#1}}}
\newcommand{\myemph}[1]{{\color{black}\emph{#1}}}
\newcommand{\probdef}[3]{\prob{\probname{#1}}{Input:}{#2}{Question:}{#3}{as}}
\newcommand{\bamex}{\textsc{BAM}}
\newcommand{\bamOPT}{\textsc{BAM}-\ensuremath{\mathsf{OPT}}}
\newcommand{\BAMproblong}{\textsc{Boolean Attribute Model}}
\newcommand{\BAMprob}{\bamex}
\newcommand{\BAMwcprob}{\textsc{BAM with Cares}}
\newcommand{\BAMwhprob}{\textsc{BAM with Has}}
\newcommand{\COLprob}{\textsc{3\nobreakdash-Coloring}}
\newcommand{\has}{\ensuremath{\mathsf{has}}} 
\newcommand{\have}{\ensuremath{\mathsf{have}}}
\newcommand{\having}{\ensuremath{\mathsf{having}}}
\newcommand{\cares}{\ensuremath{\mathsf{cares}}}
\newcommand{\care}{\ensuremath{\mathsf{care}}}
\newcommand{\bamsymb}{\ensuremath{\mathcal{M}}}
\newcommand*{\rank}[2]{r_{#1} (#2)}
\newcommand*{\score}[2]{\textup{score}_{#1} (#2)}
\newcommand{\scoreT}[3]{\textup{score}^{#1}_{#2}(#3)}

\newcommand{\todoCinline}[1]{\todo[inline, linecolor=blue!70!black, backgroundcolor=blue!10]{C: #1}}
\newcommand{\todoM}[1]{\todo[linecolor=orange!70!black, backgroundcolor=orange!10]{M: #1}}

\newcommand{\todoH}[1]{\todo[linecolor=green!70!black, backgroundcolor=green!10]{H: #1}}
\newcommand{\todoHinline}[1]{\todo[inline, linecolor=green!70!black, backgroundcolor=green!10]{H: #1}}
\newcommand{\attr}{\ensuremath{\alpha}}
\newcommand{\vertexalt}[1]{\ensuremath{c_{#1}}}

\newcommand{\dummyalt}[2]{\ensuremath{d_{#1}^{#2}}}
\newcommand{\type}{\textit{type}}

\newcommand{\aaa}{\ensuremath{\mathcal{C}}}
\newcommand{\vvv}{\ensuremath{\mathcal{V}}}
\newcommand{\RR}{\ensuremath{\mathcal{R}}}
\newcommand{\ppp}{\ensuremath{\mathcal{P}}}
\newcommand{\attrset}{\ensuremath{\mathsf{AT}}}
\newcommand{\dummyatt}[2]{\ensuremath{\delta_{#1}^{#2}}}
\newcommand{\typescore}[2]{\ensuremath{S_{#1}(#2)}}
\newcommand{\mypara}[1]{
  
  \smallskip

  \noindent\textbf{#1}}

\newcommand{\myunderline}[1]{
  \noindent\underline{#1}}

\allowdisplaybreaks
\begin{document}
\frenchspacing

\maketitle
\todoHinline{Don't we need to add addresses and email addresses as well?}
\todoCinline{I've added Vienna, but from older accepted papers it seems that most people do not give more info. I've added our email adress in general form should I add the ac ones instead?}
\begin{abstract}
  We study the computational complexity of explaining preference data through Boolean attribute models (BAMs), motivated by extensive research involving attribute models and their promise in understanding preference structure and enabling more efficient decision-making processes.
  In a BAM, each alternative \emph{has} a subset of Boolean attributes, each voter \emph{cares} about a subset of attributes,
  and voters prefer alternatives with more of their desired attributes.
  In the BAM problem, we are given a preference profile and a number~$k$, and want to know whether there is a Boolean $k$-attribute model explaining the profile.

  We establish a complexity dichotomy for the number of attributes $k$: BAM is linear-time solvable for $k \leq 2$ but NP-complete for $k \geq 3$.
  The problem remains hard even when preference orders have length two.
  On the positive side, BAM becomes fixed-parameter tractable when parameterized by the number of alternatives $m$.
  For the special case of two voters, we provide a linear-time algorithm.

We also analyze variants where partial information is given: When voter preferences over attributes are known (\BAMwcprob) or when alternative attributes are specified (\BAMwhprob), we show that for most parameters \BAMwcprob\ is more difficult whereas \BAMwhprob\ is more tractable except for being NP-hard even for one voter.
\end{abstract}

\section{Introduction}
\todoHinline{Bib entries are inconsistent! Typos in the entries as well. Please do not use ``don't'' or ``can't'' abbreviations!}
\todoCinline{Removed the occurence I found.}
\todoHinline{I now replace ``With'' with ``with'' (small letter) since it should not be capitalized.}
What patterns underlie sets of human (and other) preferences in their various applications?
Preferences in practice are not arbitrary orderings; people prefer alternatives that better satisfy their underlying criteria or possess characteristics they desire, and the distributions of such criteria and their desirability are structured.
For one example, in a speed-dating experiment,
\citet{fisman_gender_2006} found that on average the recruited women preferred intelligent partners from more affluent neighborhoods, men preferred physically attractive partners, and both men and women accepted partners from less densely populated areas.
We study the problem of how to uncover such structure from the pure revealed preferences.

A parsimonious and interpretable model that captures the above dynamic between criteria and their desirability arises from binary characteristics.
In this framework, each alternative possesses some subset of Boolean attributes (e.g., a restaurant either accepts credit cards or does not), and each individual cares about some subset of these attributes.
An individual prefers alternative $a$ to alternative~$b$ if and only if $a$ possesses more of the attributes they care about than $b$ does.
For instance, if someone cares about credit card acceptance and vegan options and restaurant $A$ has both while restaurant $B$ has only vegan options, then $A$ is preferred to~$B$.
Given a preference profile $\mathcal{P}$ (a set of preferences of voters over alternatives), we are hence interested in a \myemph{Boolean attribute model (BAM)} for $\mathcal{P}$, that is, (i) a set of attributes, (ii) an assignment \has\ of a subset of attributes to each alternative, and (iii) an assignment \cares\ of a subset of attributes to each voter such that, for each voter $v$, all of its pairwise preferences $a \succ b$ are explained by $a$ \having\ more attributes that $v$ \cares\ about than $b$.
For parsimony we aim to minimize the number of attributes in such models.
We denote by \BAMproblong\ (\BAMprob) the computational problem where we are given a preference profile and an integer $k$ and want to determine whether a Boolean attribute model with at most $k$ attributes exists.

Boolean attribute models and natural continuous variants are frequent in the literature (see~\citet{fisman_gender_2006,bhatnagar2008sampling,kunnemannSubquadraticAlgorithmsSuccinct2019a,chengStableMatchingsRestricted2023} for just a few examples).
However, we are not aware of work that considers \myemph{computing} such attribute models (and checking how well they fit the data).
Our goal here is to contribute towards this direction by studying the computational complexity of this problem, with the goal of identifying hard and tractable cases and thereby informing practical algorithm design.

\smallskip%
Obtaining (Boolean) attribute models confers several benefits.
First, it is intrinsically interesting to check whether practical preference profiles have attribute models with few attributes as this would contribute to our understanding of (human) preference formation and decision making. However, we currently lack the necessary algorithmic research.

Second, knowing that an application allows for parsimonious attribute models allows us to restrict the domains of preferences that we consider when understanding properties of decision-making processes and designing algorithms.
This enables more scalable algorithms: For instance, computing stable matchings generally needs quadratic time in the number~$n$ of agents but if a Boolean $k$-attribute model is known, then they can be computed in $O(4^k \cdot n \cdot (k + \log n))$ time, that is, almost linear time for small~$k$~\cite{kunnemannSubquadraticAlgorithmsSuccinct2019a}.
Third, if we know that certain applications admit small attribute models, then preference learning~\cite{FurnkranzH10,FurnkranzH17}, more specifically eliciting preferences from voters, becomes more tractable: Instead of having to ask voters for rankings of the (potentially many) alternatives, we can ask them which (of the few) attributes are important to them.
For instance, \citet{BenabbouDPV16} provide an application of this principle to compute Borda winners efficiently.
See also \citet{FefferSLH23} for a recent survey of preference elicitation in the context of participatory machine learning.

\newcommand{\paranph}{NPh}
\newcommand{\npc}{NPc}
\newcommand{\fpt}{FPT}
\newcommand{\paracomment}[1]{{\small(#1)}}
  \begin{table}[t]
  \caption{Result overview; $n$ is the number of voters, $m$ the number of alternatives, and $k$ an upper bound on the number of attributes. ``-'' means the \fpt\ result follows from the result for the single parameter~$m$. ``\paranph~($p\ge c$)'' means that the corresponding problem remains NP-hard even if the parameter~$p$ has constant value~$c$.}
  \centering 
\resizebox{\columnwidth}{!}
{
  \begin{tabular}{@{}r c @{\;\,} l @{\quad} c @{\;\,} l @{\qquad} c@{\;\,} l@{}}
    \toprule 
    Parameter   & \multicolumn{2}{c}{\BAMprob} 
    & \multicolumn{2}{c}{\BAMwcprob} 
    & \multicolumn{2}{c}{\BAMwhprob}\\\midrule
    In general& \npc & [T\ref{thm:BAMNPklistlength}] & \quad\npc{} & [T\ref{thm:BAMwcNPh}] & \qquad\npc{} & [T\ref{thm:BAMwhNPh}]\\\midrule
    {$n$} & ? & & \quad? &  & \qquad\paranph & [T\ref{thm:BAMwhNPh}] \\
                & \multicolumn{2}{c}{\small $O(m)$ if $n = 2$} & &  & \multicolumn{2}{c}{\paracomment{$n\geq1$}}  \\[1ex]
{$m$} & \fpt & [T\ref{thm:BAMm}] & \quad\paranph{~}\paracomment{$m\geq 3$}& [T\ref{thm:BAMwcNPh}] & \qquad\fpt{} & [T\ref{thm:BAMwhm}]\\
[1ex]
$k$ & \paranph{}& [T\ref{thm:BAMNPklistlength}] & \quad\paranph{}~\paracomment{$k\geq6$} & [T\ref{thm:BAMwck}] & \qquad \fpt\ & [T\ref{thm:BAMwhk}]\\
\midrule

    $n+m$ & -- &-- & \quad \fpt\ & [T\ref{thm:BAMwcnm}] & \qquad-- &-- \\
$n+k$ & \fpt\ & [C\ref{cor:BAMnk}] & \quad \fpt\ & [C\ref{thm:BAMwcnk}] & \qquad-- &--\\
$m+k$ & -- &-- & \quad \fpt\ & [T\ref{thm:BAMwcmk}] & \qquad-- &--\\
    \bottomrule
  \end{tabular}}
  \label{tab:results}
\end{table}

\smallskip
\paragraph{Our contribution.}
In general, \BAMprob\ is NP-hard, which we show via a reduction from computing proper graph colorings.
Given this general hardness, we (a) investigate more closely what parameters $p \in \mathbb{N}$ of the input make the problem hard or tractable and (b) check how additional information given in the input affects the complexity.
An overview of most of our results is given in \cref{tab:results}.

As to (a), we obtain either the favorable fixed-parameter tractability (FPT), meaning that the problem is solvable in $f(p) \cdot |I|^{O(1)}$ time, where $|I|$ is the input size. In other words, there are efficient algorithms for small parameter values.
Or we obtain NP-hardness even for constant values of the parameter $p$.
As parameters $p$ we start with the natural candidates: the number $n$ of voters, the number $m$ of alternatives, and the number $k$ of attributes.
For $k$, we give a dichotomy for \BAMprob, showing NP-hardness for computing Boolean 3\nobreakdash-attribute models (even for preference orders of length at most two), whereas for two attributes the problem is polynomial-time solvable via a reduction to $2$-SAT.
For parameter $m$, the problem turns out to be FPT.
For $n = 2$ voters we show that the problem can be solved in linear time by deriving lower bounds on the needed number of attributes and showing that there is always an optimal solution that meets these lower bounds exactly.
We leave the complexity wrt.\ parameter $n$ as an open question, but when combining $n$ with $k$ we obtain FPT.
All these results are given in \cref{sec:BAM}.

As to (b), in some applications BAMs are partially given already, that is, we may already know (i) which attributes voters care about, leading to the problem \BAMwcprob\ or (ii) which attributes the alternatives have, corresponding to problem \BAMwhprob.
We show that also both these problems are NP-hard and we analogously subject them to a parameterized complexity analysis.
Here, generally \BAMwcprob\ turns out to be harder than \BAMprob\ (see \cref{sec:BAMwc}) whereas \BAMwhprob\ is easier, up to the parameter $n$, where \BAMwhprob\ is NP-hard already for one voter (see \cref{sec:BAMwh}).

\paragraph{Related work.}
Attribute models of preferences are fundamental and widely used in decision theory~\cite{yu_multiplecriteria_1985,Siskos2016}, social choice theory~\cite{LangX16}, and machine learning~\cite{Labreuche11}.
Attribute models have also been used in McFadden's conditional logit model to statistically predict discrete choices of individuals~\cite{McFadden1974}
and as a decision-making heuristic of Tversky's elimination by aspects~\cite{Tversky1972}. %
Often the attributes involve larger domains but Boolean domains are simple, interpretable, and cover relevant basic cases~\cite{LangX09}.

To relate to other works below it is useful to consider the following alternative view of BAMs.
Implicitly, in a $k$-attribute BAM we define a utility function for each voter:
We associate each alternative $a$ (resp.\ each voter~$v$) with a vector $\has(a) \in \{0, 1\}^k$ (resp.\ a vector $\cares(v) \in \{0, 1\}^k$).
The utility of $v$ for~$a$ is the scalar product of $\has(a)$ and $\cares(v)$, i.e., the weighted sum $\sum_{i \in [k]} \has(a)[i] \cdot \cares(v)[i]$.
The so-defined utilities must rise strictly with voter's preferences.
Thus, having a $k$-attribute BAM amounts to learning, for each voter, a utility function that is the sum of utilities for the individual attributes.

A further motivation of BAMs comes from explaining AI systems' decisions~\cite{HolzingerSMBS20}: Here, the input preferences may stem from black-box machine-learning models.
We may even know the attributes (features) that each of the alternatives have and were taken into consideration by the models.
The task is then to select the attributes that the models relied on in their decision, that is, what attributes they cared about~\cite{Labreuche11}.

A similar set of tasks stems from multi-criteria decision making~\cite{yu_multiplecriteria_1985} and preference learning.
Herein, one or multiple~\cite{AuriauBMM24} decision makers reveal their preferences and we try to learn their utility functions. %
So-called UTA methods (UTilit\'{e}s Additives or additive utilities in English)~\cite{Siskos2016} are popular for this task: We assume that the utility functions are weighted sums over the attributes and we try to learn the underlying attributes and the utility gained from each attribute.
In a setting with multiple decision makers we additionally want to cluster the input preferences and learn a utility function for each decision maker over his corresponding cluster~\cite{AuriauBMM24}.
In BAMs we also try to learn simple additive utility functions from revealed preferences but we do not cluster the preferences directly (although a clustering may be derived from voters with similar \cares-mappings).

In social choice theory, voting in combinatorial domains~\cite{LangX16} is loosely related to BAMs. Here, we consider the alternatives to be tuples of attribute values; often the attributes are Boolean, corresponding to multiple-issue voting.
The voters express (usually complete) preferences over alternatives. %
Generally, straightforward issue-wise (attribute-wise) voting rules then lead to undesirable outcomes and one tries to find better voting rules.
In contrast, in BAMs we assume that voters can only approve or disapprove of each attribute and we may have incomplete preferences.
Crucially though, in combinatorial voting it is usually assumed that an attribute model is given whereas we want to compute such models.

Finally, preferences with (Boolean) few-attribute models form a restricted preference domain.
Such domains and how they can allow for more efficient decision-making procedures are a mainstay of social choice theory~\cite{ElkindLP22}.
The group-separable preference domain~\cite{Inada1964,Inada1969} could be interpreted as having a sequence of binary attributes with exponentially decreasing weights and voters evaluate attributes positively or negatively.   
More closely related are the geometric preference under~$\ell_p$ norm~\cite{chenOnedimensionalEuclideanDomain2017,petersRecognisingMultidimensionalEuclidean,Chen2022Manhattan} where the voters and alternatives are located in a geometric space and
voters prefer an alternative that is closer to him (under $\ell_p$ norm).
Another popular domain restriction are so-called single-peaked preferences.
However, there is little evidence that real-world preferences are single-peaked~\cite{SuiFB13,Przedmojski16} and thus researchers called for investigation into multi-(but low-)-dimensional variants in particular of single-peaked domains~\cite{BarberaGS93}.
In a sense BAMs constitute one of the simplest multi-dimensional restricted preference domains and thus may inform investigation into more sophisticated variants.

\section{Preliminaries}\label{sec:prelim}
\appendixsection{sec:prelim}
\todoHinline{The following sentence should be updated for the AAAI camera.}
\todoCinline{Left for now as it is needed for arxiv.}
(Full) proofs for results marked by (\appsymb) are deferred to the supplementary material.
In this section, we present necessary concepts for our Boolean attribute model and collect some fundamental properties thereof. %
Given a non-negative integer~$t\in \mathbb{N}$, let $[t]$ denote the set~$\{1,\ldots, t\}$.
We assume basic knowledge of parameterized complexity and refer to the textbook by \citet{CyFoKoLoMaPiPiSa2015} for more details.
\paragraph{Preference profiles and BAMs.}
A \myemph{preference profile} (\myemph{profile} in short) is a triple~$\ppp=(\aaa, \vvv, \RR)$, consisting of a set~$\aaa$ of $m$ alternatives, a set~$\vvv=\{v_1,\ldots, v_n\}$ of $n$ voters, and a collection~$\RR=(\succ_{v_1},\ldots, \succ_{v_n})$ of (possibly incomplete) \myemph{strict preference orders} such that each~$\succ_i$, is a linear order of a subset of $\aaa$
and represents the preferences (aka.\ ranking) of voter~$v_i$ over some subset of~$\aaa$, $i\in [n]$.
For instance, for $\aaa=\{1,2,3,4\}$, an incomplete preference order of a voter~$v$ can be $3\succ 1 \succ 2$; we omit the subscript if it is clear from the context which voter we refer to. 
This means that voter~$v$ prefers $3$ to $1$, and $1$ to $2$.
Alternative~$4$ is not ranked in his preference order.
The \myemph{rank} of an alternative~$c$ for a voter $v$ is the number of alternatives that $v$ prefers to~$c$,
i.e., $\rank{v}{c}\coloneqq |\{b \mid b \succ_v c \}|$.
$\rank{v}{c}$ is undefined if $v$'s preference order does not rank~$c$.
The \myemph{length} $|\succ_v|$ of $v$'s preference order is the number of alternatives in $\succ_v$.
For $3\succ 1 \succ 2$, we infer that $\rank{v}{2}=2$ and $|\succ|=3$. 

A \myemph{Boolean attribute model} for a profile $\ppp=(\aaa, \vvv, \RR)$ specifies for each alternative which attributes it \myemph{\has} and for each voter which attributes he \myemph{\cares} about.
More specifically, let $\bamsymb=(\attrset, \has, \cares)$ be a triple, where $\attrset$ denotes a set of attributes, and $\has\colon \aaa \to 2^{\attrset}$ and $\cares\colon \vvv \to2^{\attrset}$ two attribute functions for the alternatives and voters, respectively.
We say that an alternative~$c\in \aaa$ $\has$ an attribute~$\attr$ if $\attr \in \has(c)$ (or \have, depending on grammatical necessity), and that $c$ $\has$ $\ell$ attributes if $|\has{}(c)| = \ell$.
A similar convention applies for the use of \textbf{cares}.

The \myemph{score} of a voter~$v$ for an alternative~$c$ under~$\bamsymb$ is defined as
  $\scoreT{\bamsymb}{v}{c} \coloneqq |\has{(c)} \cap \cares{(v)}|$. 
We omit the superscript~$\bamsymb$ from the score if it is clear from the context which~$\bamsymb$ we refer to.
We say that~$\bamsymb$ \myemph{explains} the preference order of voter~$v\in \vvv$ if
for each two alternatives~$c$ and $d$  with $c \succ_v d$
it holds that $\score{v}{c} > \score{v}{d}$.
 Note that we do not require the backward implication to hold since not every alternative is ranked.
 Accordingly, we say that $\bamsymb$ is a \myemph{Boolean attribute model} (\myemph{BAM} for short) for~$\ppp$ if it explains the preference order of every voter. 
 
 Intuitively, the model aims to explain the voters' preferences by assigning each voter a set of attributes they \care\ about, and assigning each alternative a set of attributes it possesses.
 Voters then prefer alternatives that possess more attributes they \care\ about.
 If $|\attrset|=k$, then $\bamsymb$ is also referred to as a \myemph{$k$-Boolean attribute model} (\myemph{$k$-BAM}).
 
 \begin{remark}
   For brevity's sake, we sometimes also use the following vector representation for a $k$-BAM~$\bamsymb$. 
   The $\has\colon \aaa \to \{0,1\}^k$ (resp.\ $\cares\colon \vvv \to \{0, 1\}^k$) function defines for each alternative~$c\in \aaa$ (resp.\ each voter~$v\in \vvv$) a binary vector of length~$k$, where the attribute set is simply~$[k]$ such that a $1$ at coordinate~$z$ means the alternative \has\ (resp.\ the voter \cares\ about) attribute~$z$. 
  
   The score of alternative~$c\in \aaa$ by a voter~$v\in \vvv$ is given by the scalar product:
   $\scoreT{\bamsymb}{v}{c}\coloneqq \has{(c)} \cdot \cares{(v)}$.
  \end{remark}

Generally, our goal is to find a BAM that explains the voters' preferences with as few attributes as possible.
We define the three decision problems that form the core of our work:
\probdef{\bamex}
{A profile~$\ppp$ and a non-negative integer $k$.}{Is there a $k$-Boolean attribute model for \(\mathcal P\)?}

\probdef{\BAMwcprob\ (\text{resp.} \BAMwhprob)}
  {A preference profile~$\ppp = (\aaa, \vvv, \RR)$ and a function~$\cares\colon \vvv \to 2^{[k]}$
    (resp.\ $\has\colon \aaa \to 2^{[k]}$).}
  {Is there a function~$\has\colon \aaa \to 2^{[k]}$
     (resp.\ $\cares\colon \aaa \to 2^{[k]}$)
    s.t.\ $(\has, \cares)$ is a $k$-BAM for $\ppp$?}

Clearly, our central problems are contained in NP:
\begin{observation}\label{obs:np-contain}
    Checking whether $(\attrset,\has,\cares)$ explains a profile $\ppp = (\aaa,\vvv,\RR)$ is doable in polynomial time. %
\end{observation}
This follows from the fact that one can determine the score of an alternative for a voter in polynomial time, and then iterate over every voter and every pair of alternatives to check if the voter's preference order is explained.
\toappendix{We can also define the optimization-variant of our problem:
\probdef{\bamOPT}
{A profile~$\ppp$.}{What is the smallest $k\in\mathbb{N}$ s.t. there exists a $k$-Boolean attribute model for \(\mathcal P\)?}
Note that for \BAMwcprob\ and \BAMwhprob\ an optimization variant cannot be defined analogously, as the number of attributes is implicitly given through the $\cares$-function, respectively, the $\has$-function.
}
\paragraph{Fundamental properties.} We now consider structural properties of BAMs.
First, we provide some bounds on the number of attributes cared about by a voter (resp.\  possessed by an alternative).
\begin{restatable}[\appsymb]{lemma}{lempreflistlength}\label{lem:attributes-bound}
  For each $k$-BAM that explains a preference profile $\ppp=(\aaa,\vvv,\RR)$, the following holds.
  \begin{compactenum}[(i)]
    \item\label{care-bound} For all~$v\in \vvv$, it holds that $|\succ_v|-1 \le$ $|\cares{(v)}|$ $\le k$.
    \item\label{has-bound} For all~$v\in \vvv$ and all~$c$ in $\succ_v$, it holds that
    $|\succ_v| - \rank{v}{c} - 1 \le |\has{(c)}|\le k - \rank v c$.
  \end{compactenum}
\end{restatable} 
\appendixproofwithstatement{lem:attributes-bound}{\lempreflistlength*}{
  \begin{proof}
    \begin{compactenum}[(i)]
    \item  The second part of the inequality, $k\geq|\cares{(v)}|$, holds trivially, as $\cares(v)\subseteq\attrset$ and $|\attrset|=k$. It therefore remains to show the second part of the inequality, $|\cares{(v)}| \ge |\succ_v|-1$.
   As the scores of alternatives along a preference order must be strictly decreasing, if a preference order has $|\succ_v|$ many ranked alternatives, the list must contain at least $|\succ_v|-1$ scores. If $|\cares{(v)}| < |\succ_v|-1$, then
   there can be at most $|\succ_v|-1$ scores, as the maximum achievable score is $|\succ_v| - 2$ and there are $|\succ_v|-1$ values in the set \([|\succ_v|-2]\). 
   \item If an alternative is preferred to $\ell$ others in a preference order, it $\has$ at least $\ell$ attributes, and if it is ranked worse than $\ell$ others, it $\has$ at most $k-\ell$ attributes.
  \end{compactenum}
\end{proof}}
\toappendix{
  Note that the lower bounds on the number of attributes that a voter \cares\ about and an alternative \has\ derived from \cref{lem:attributes-bound} are not always sufficient. 
  We give an example to illustrate.
\begin{example}
Consider a profile with four alternatives and three voters, where the preference orders are as follows:
\begin{align*}
v_1&\colon c_1\succ c_2\succ c_3\succ c_4, \quad v_2 \colon c_4\succ c_3\succ c_2\succ c_1,\\
v_3&\colon c_4\succ c_1\succ c_2. 
\end{align*}
We claim that the smallest possible number of attributes for this profile is $k=6$.
Observe that $v_1$ and $v_2$'s preference orders are reverse to each other.
To explain voter~$v_1$'s preference order, it must hold that $|\has(c_1)\setminus\has(c_4)|\geq 3$,
and to explain~$v_2$' preference order, it must hold that $|\has(c_4)\setminus\has(c_1)|\geq3$.
A $6$-BAM can be constructed by creating two disjoint sets of attributes~$\attrset_1$ and $\attrset_2$ of equal size~$3$.
Voter $v_1$ only \cares\ about $\attrset_1$ and voter~$v_2$ only $\attrset_2$.
Alternative $c_1$ \has\ all attributes from $\attrset_1$ and none of $\attrset_2$.
Alternative $c_2$ \has\ all but one attribute from $\attrset_1$ and one of $\attrset_2$.
This construction continues for $c_3$ and $c_4$ analogously.

Note that this is the only way to construct a $6$-BAM due to the preference orders of the first two voters: We must have three attributes in $\has(c_1)\setminus \has(c_4)$ and three other attributes in~$\has(c_4)\setminus \has(c_1)$.
Now, since $|\has(c_2)|=2+1>\max_{v\in V}|\succ_v| - \rank{v}{c_2} - 1=4-1-1=2$, it follows that the lower bound of $\has(c)$ from \cref{lem:attributes-bound}(\ref{has-bound}) does not always suffice to explain a profile using a minimum amount of attributes.

 Voter $v_3$ then \cares\ about at least one attribute from $\has(c_1)\setminus\has(c_2)$ and at least two attributes from $\has(c_4)\setminus\has(c_2)$. As $\has(c_4)\cap\has(c_1)=\emptyset$, it follows that $\cares(v_3)\geq 3> |\succ_{v_3}|-1=2$ and therefore the lower bound of $\cares(v)$ from \cref{lem:attributes-bound}(\ref{care-bound}) does not always suffice to explain a profile using a minimum amount of attributes.
\end{example}}
We can also observe that each $k$-BAM must be able to explain the difference in ranks of an alternative in two preference orders.
This leads to the following lemma.
\begin{restatable}[\appsymb]{lemma}{lemrankdist}\label{lem:rank-dist}
    If $\ppp = (\aaa,\vvv,\RR)$ admits a $k$-BAM, then for all $c \in \aaa$, and $v, w \in \vvv$ with $\rank v c \ge \rank w c$, it holds that    
    $ |\succ_w| - \rank{w}{c} + \rank{v}{c} \le   k + 1$.
\end{restatable} 
\appendixproofwithstatement{lem:rank-dist}{\lemrankdist*}{
\begin{proof}
  Recall that $\rank{v}{c} \ge \rank{w}{c}$.
  Applying \cref{lem:attributes-bound}\eqref{has-bound} for $v$, we infer that 
  \begin{equation*}
    k - \rank v c \ge |\has{(c)}|
  \end{equation*}
  and for $w$
  \begin{equation*}
    |\has{(c)}| \ge |\succ_w| - \rank w c - 1.
  \end{equation*}
  from which we obtain
  \begin{align*}
    k - \rank v c & \ge |\succ_w| - \rank w c- 1 \iff
    \\k + 1 &\ge |\succ_w| - \rank w c + \rank v c.\qedhere
  \end{align*}
\end{proof}}

\begin{example}
  We examine each pair of occurrences of an alternative in pairs of voters, and derive a lower bound for $k$. Consider the following example, with voters $v$ and $w$:
\begin{align*}
    v\colon a \succ b \succ c \succ d \succ e,\text{ and }
    w\colon f \succ d \succ g \succ h \succ i.
\end{align*}
For alternative $d$, we have $\rank w d = 1$ and $\rank v d = 3$.
Lemma \ref{lem:rank-dist} implies that $k \geq |\succ_w|-1+3-1=6$.
Intuitively, alternative~$d$ needs to $\have$ at least three attributes in order to explain voter~$w$ (since it is ranked higher than three other alternatives by~$w$),
but there should also be at least three attributes that $d$ does not $\have$ in order to explain voter $v$ (since three other alternatives are ranked higher than~$d$). %
\end{example}
\toappendix{
It may seem intuitive that $k=\displaystyle\max_{c \in \aaa, v,w\in\vvv}|\succ_w| - \rank{w}{c} + \rank{v}{c}-1$ attributes suffice to explain every profile~$\ppp$. We provide an example to show that this is not the case:
\begin{example}
Consider a profile~$\ppp$ with three alternatives and three voters, where the preference orders are as follows:
\begin{align*}
v_1\colon& c_1\succ c_2,\quad  v_2\colon  c_2\succ c_3,\quad v_3\colon c_3\succ c_1.
\end{align*}
For this profile, we infer by \cref{lem:rank-dist} that $k \ge \displaystyle\max_{c \in \aaa, v,w\in\vvv}|\succ_w| - \rank{w}{c} + \rank{v}{c}-1=2$.
However, we claim that every BAM for $\ppp$ has at least $3>k$ attributes.
First, we observe that for all alternatives $c\in\{c_1,c_2,c_3\}$, it must hold that $\has(c)\geq 1$ since each alternative is ranked higher by at least one other alternative.
For $k=2$, it must hold that $\has(c)\leq 1$ for all alternatives~$c$ since
every alternative is ranked lower than at least one other alternative. 
Therefore, $\has(c)=1$ holds for all alternatives~$c$. 
As each pair of alternatives is compared it follows that no two alternatives can \have\ the same subset of attributes.
For three alternatives, we then need at least three attributes. 
A $3$-BAM that explains \ppp\ is $\has(c_1)=\cares(v_1)=\{\attr_1\}$, $\has(c_2)=\cares(v_2)=\{\attr_2\}$, and $\has(c_3)=\cares(v_3)=\{\attr_3\}$.
\end{example}
}
Finally, we show that a $k$-BAM always exists for large enough~$k$.
\begin{restatable}[\appsymb]{lemma}{lemtrivupperbound}\label{lem:triv-upper-bound}
  For a profile with $m$ alternatives and $n$ voters, a $k$-BAM with $k\ge (m-1)\cdot m$ or $k \ge (m-1)\cdot n$ always exists.
\end{restatable}
\appendixproofwithstatement{lem:triv-upper-bound}{\lemtrivupperbound*}{
  \begin{proof}
    We show the statement for $k=(m-1)\cdot m$ constructively:
    Let each alternative $c$ \have\ $m-1$ attributes $\attr_{c}^1, \dots, \attr_{c}^{m-1}$ that only $c$ $\has$, i.e.,
    \begin{equation*}
      A \coloneqq \{\attr_{c}^i \mid c \in C, i \in [m - 1]\}
    \end{equation*}
    such that for all $a, c \in \aaa$,
    \begin{equation*}
      \attr_a^i \in \has{}(c) \iff a = c \wedge 1 \le i \le m-1.
    \end{equation*}
    Then, let every voter $v \in \vvv$ \textbf{care} about as many of the attributes of any given alternative as they need.
    This can be described formally by
    \begin{equation*}
      \cares{}(v) \coloneqq \{\attr_{c}^i \mid v \text{ ranks } c \text{ and } 1 \le i \le |\succ_v| - \rank{v}{c}-1\}.
    \end{equation*}
    The above is well-defined, as each voter needs to $\care$ about at most $(m - 1)$ many private attributes of a given alternative, because a list of $m$ alternatives is can be described by $m - 1$ attributes.
    
    We claim that the functions $(\has{}, \cares{})$ constitute a BAM.
    For every $v \in \vvv$, $c \in \aaa$, if $v$ ranks $c$, the score obtained by $c$ for $v$ equals $|\succ_v| - \rank{v}{c}$, which strictly decreases when traversing down the preference order and is non-negative by the definition of the rank.
    
    The analogous statement for $k = (m-1)\cdot n$ can be shown similarly by, instead of introducing $(m-1)$ private attributes for each alternative, introducing $(m-1)$ private attributes for each voter.
  \end{proof}}

\begin{remark}
  \cref{lem:triv-upper-bound} immediately implies fixed-parameter tractability for $k$-BAM wrt.~$m$ since a profile with $m$ alternatives has at most $O(2^m\cdot m!)$ different (incomplete) preference orders.

  Moreover, for profiles with complete preferences,
  we have by \cref{lem:attributes-bound}\eqref{care-bound} and \cref{lem:triv-upper-bound} that $m-1 \le k \le m(m-1)$. %
  Therefore, the parameters $k$ and $m$ are equivalent from the parameterized perspective.
  In other words, a voting problem is FPT wrt.\ $m$ if and only if it is FPT wrt.\ $k$. %
\end{remark}

\todoM{Revisit}

\section{BAM}\label{sec:BAM}
\appendixsection{sec:BAM}
In this section, we consider the first computational problem~\BAMprob{}.
We first establish NP-completeness and then propose some parameterized and polynomial-time algorithms for some special cases.

\paragraph{General complexity.}
\BAMprob{} is NP-hard by a reduction from the NP-complete problem below~\cite{papadimitriouComputationalComplexity1994}.

\probdef{3-Coloring}
{An undirected graph~$G = (U, E)$.}
{Does $G$ admit a \myemph{proper $3$-coloring}, i.e., a function~$\chi\colon U \to [3]$
  s.t.\ no two adjacent vertices have the same value?}

\begin{restatable}[\appsymb]{theorem}{thmBAMNPklistlength}\label{thm:BAMNPklistlength}
  \bamex{} is NP-complete; it remains NP-hard even if $k = 3$ and every preference order has length~two.
\end{restatable} 
{
\begin{proof}
  NP-membership follows directly from \cref{obs:np-contain}.
  To show NP-hardness, we reduce from \COLprob{}, which results in an instance of \bamex\ with $k=3$
  such that every preference order has length two.

  Let $G=(U, E)$ be an instance of \COLprob{}.
  W.l.o.g, assume that no vertices have degree zero. We create an instance $I'=(\ppp=(\aaa,\vvv,\RR),k=3)$ as follows.

  \mypara{Alternatives.} For each vertex~$u\in U$, we add an alternative $\vertexalt{u}$ to $\aaa$.
  Additionally, we add to $\aaa$ three groups of dummy alternatives containing in total seven dummies
  $D_1 \coloneqq \{\dummyalt{1}{1}, \dummyalt{1}{2}, \dummyalt{1}{3}\}$,
  $D_2 \coloneqq \{\dummyalt{2}{1}, \dummyalt{2}{2}, \dummyalt{2}{3}\}$,
  and $D_3\coloneqq \{\dummyalt{3}{}\}$.
  We will ensure that the alternatives in $D_1$ will each \have\ a distinct attribute,
  the alternatives in $D_2$ will each \have\ a distinct pair of attributes,
  while the single alternative in~$D_3$ will \have\ all three attributes.

  \mypara{Voters and their preferences.} 
  \begin{compactitem}[--]
    \item[--] For each vertex~$u\in U$, we add three voters $v_{u}^1,v_{u}^2,v_{u}^3$ with preference orders:
    $v_{u}^1\colon \dummyalt{2}{1} \succ \vertexalt{u}$,
    $v_{u}^2\colon \dummyalt{2}{2} \succ \vertexalt{u}$,
    and $v_{u}^3\colon \dummyalt{2}{3} \succ \vertexalt{u}$ (ensuring $\vertexalt{u}$ \has\ at most one attribute).
    \item[--] For each edge~$\{u,w\}\in E$, we add two voters~$v_{u,w}$ and $v_{w,u}$ with preference orders
    $v_{(u,w)}\colon \vertexalt{u} \succ \vertexalt{w}$ and
    $v_{(w,u)}\colon \vertexalt{w} \succ \vertexalt{u}$. This will ensure simultaneously that $\vertexalt{u}, \vertexalt{w}$ \have\ exactly one attribute and they are distinct.

    \item[--] We add $24$ unnamed dummy voters in four groups~$V_1,V_2,V_3, V_4$. These voters ensure that the dummy alternatives \have\ the desired number of attributes as mentioned before. 
    The first group~$V_1$ has six voters with preference orders:
    $\dummyalt{1}{1} \succ \dummyalt{1}{2}$,
    $\dummyalt{1}{2} \succ \dummyalt{1}{1}$,
    $\dummyalt{1}{1} \succ \dummyalt{1}{3}$,
    $\dummyalt{1}{3} \succ \dummyalt{1}{1}$,
    $\dummyalt{1}{2} \succ \dummyalt{1}{3}$,
    $\dummyalt{1}{3} \succ \dummyalt{1}{2}$.

    The second group~$V_2$ has $9$ voters with preference orders:
    $\dummyalt{2}{1} \succ \dummyalt{1}{1}$,
    $\dummyalt{2}{1} \succ \dummyalt{1}{2}$,
    $\dummyalt{2}{1} \succ \dummyalt{1}{3}$,
    $\dummyalt{2}{2} \succ \dummyalt{1}{1}$,
    $\dummyalt{2}{2} \succ \dummyalt{1}{2}$,
    $\dummyalt{2}{2} \succ \dummyalt{1}{3}$,
    $\dummyalt{2}{3} \succ \dummyalt{1}{1}$,
    $\dummyalt{2}{3} \succ \dummyalt{1}{2}$,
    $\dummyalt{2}{3} \succ \dummyalt{1}{3}$.

    The third group~$V_3$ has $6$ voters with preference orders:
    $\dummyalt{2}{1} \succ \dummyalt{2}{2}$,
    $\dummyalt{2}{2} \succ \dummyalt{2}{1}$,
    $\dummyalt{2}{1} \succ \dummyalt{2}{3}$,
    $\dummyalt{2}{3} \succ \dummyalt{2}{1}$,
    $\dummyalt{2}{2} \succ \dummyalt{2}{3}$,
    $\dummyalt{2}{3} \succ \dummyalt{2}{2}$.

    The last group~$V_4$ has $3$ voters with preference orders:
    $\dummyalt{3}{} \succ \dummyalt{2}{1}$,
    $\dummyalt{3}{} \succ \dummyalt{2}{2}$,
    $\dummyalt{3}{} \succ \dummyalt{2}{3}$,
  \end{compactitem}
  This concludes the construction, which can clearly be done in polynomial time. 
  It remains to show the correctness, i.e., $G$ has a proper $3$-coloring if and only if the constructed profile~$\ppp$ admits a $3$-BAM.
  \toappendixproofcontinued{%
    For the ``only if'' part, let $\chi\colon U\to[3]$ be a proper $3$-coloring.
    It is straightforward to check that the following functions~$\has\colon \aaa\to [3]$ and $\cares\colon \vvv \to [3]$ explain our profile.
     \begin{compactitem}[--]
    \item For each vertex~$u \in U$, let $\has(\vertexalt{u})\coloneqq\{\chi(u)\}$.
    \item For each~$j\in[3]$, let $\has(\dummyalt{1}{j})\coloneqq\{j\}$ and $\has(\dummyalt{2}{j})\coloneqq[3]\setminus\{j\}$. Let $\has(\dummyalt{3}{})\coloneqq[3]$. 
    \item For each dummy voter $v\in \cup_{i\in [4]}V_i$, let $x_v$ be the first-ranked alternative of the voter and $\cares(v)\coloneqq \has(x_v)$.
    \item For each vertex~$u\in U$, let $\cares(v_{u}^1)=\cares(v_{u}^2)=\cares(v_{u}^3)\coloneqq [3]$.
    \item For each edge~$\{u,w\} \in E$, let $\cares(v_{(u,w)})\coloneqq \{\chi(u)\}$ and $\cares(v_{(w,u)})\coloneqq \{\chi(w)\}$.
  \end{compactitem}
  The details are deferred to~\cref{proof:thm:BAMNPklistlength} in the supplementary material.\todoH{Check all such phrasing and replace to refer to the arXiv version.}
  
  For the ``if'' part, let $\bamsymb=(\has, \cares)$ be a $3$-BAM for $\ppp$.
  We can show that the \has\ function restricted to the ``vertex'' alternatives yields a proper $3$-coloring by observing that no two dummy alternatives have the same subset of attributes and hence every vertex has a single distinct attribute. 
  The proof is also deferred to the supplementary material.
  }{thm:BAMNPklistlength}{\thmBAMNPklistlength*}{
    We continue to show that the defined model $(\has, \cares)$ is a $3$-BAM.
    It is straightforward to verify that the preference orders of all dummy voters are explained.
    For each vertex~$u\in U$, the preference orders of the three corresponding voters $v_{u}^1$, $v_{u}^2$, and $v_{u}^3$ are explained
    because the first-ranked alternative of each of these voters has score two, whereas $\vertexalt{u}$ has score one.
    Finally, for each edge~$\{u,w\}\in E$, the preference orders of the two corresponding voters~$v_{u,w}$ and $v_{w,u}$ are explained as well because $\has(\vertexalt{u})=\cares(v_{u,w})=\{\chi(u)\}$ and $\has(\vertexalt{w})=\cares(v_{w,u})=\{\chi(w)\}$ and $\chi(u)\neq \chi(w)$. 
    This concludes the proof.
    
  For the ``if'' part, let $\bamsymb=(\has, \cares)$ be a $3$-BAM for $\ppp$. %
  We aim to show that the \has\ function restricted to the ``vertex'' alternatives yields a proper $3$-coloring. 
  To this end, let us observe that the dummy alternatives and the ``vertex'' alternative \have\ the desired number of attributes, respectively.
  \begin{restatable}[\appsymb]{claim}{claimthreecolordummies}\label{claim:3colordummies}
    \begin{compactenum}[(i)]
      \item\label{th1:dummy1} $|\has(\dummyalt{1}{z})|=1$ holds for all $z\in [3]$.
      \item\label{th1:dummy2} $|\has(\dummyalt{2}{z})|=2$ holds for all $z\in [3]$.
      \item\label{th1:dummy3} $\has(\dummyalt{3}{z})=[3]$.
      \item\label{th1:dummy4} No two dummy alternatives \have\ the same set of attributes.
      \item\label{th1:vertices} $|\has(\vertexalt{u})| = 1$ holds for all vertices~$u\in U$.
    \end{compactenum} 
  \end{restatable}
  {\begin{claimproof}{claim:3colordummies}
    The first four statements are straightforward to verify
    by checking the three groups of dummy voters sequentially. 
    As for the last statement, let us consider an arbitrary vertex~$u\in U$.
    We need to show that $|\has(\vertexalt{u})| = 1$.
    Clearly, $|\has(\vertexalt{u})| \neq 0$ since $u$ is adjacent to at least one other vertex, say~$w$, and the preference order of voter~$v_{(u,w)}$ ensures that $\vertexalt{u}$ must \have\ at least one attribute.
    Further, $\has(\vertexalt{u})$ cannot \have\ two or more attributes since for each dummy alternative~$\dummyalt{2}{z}$, $z\in [3]$, $\dummyalt{2}{z}$ \has\ exactly two attributes~(see Statement~\eqref{th1:dummy2}) and there exists a voter that prefers~$\dummyalt{2}{z}$ to~$\vertexalt{u}$.
  \end{claimproof}
  }
  By \cref{claim:3colordummies}\eqref{th1:vertices},
  we define the following function~$\chi\colon U \to [3]$ with $\chi(u)=z$ if and only if $\has(\vertexalt{u}) =\{z\}$. Clearly, this is a $3$-coloring since $\bamsymb$ is a $3$-BAM.
  Suppose, for the sake of contradiction, that $\chi$ is not a proper coloring, i.e., there exist two vertices~$u,w\in U$ such that $\{u,w\}\in E$ and $\chi(u)=\chi(w)$.
  This directly contradicts $\bamsymb$ being a $3$-BAM since voter~$v_{(u,w)}$'s preference order is not explained since he prefers~$\vertexalt{u}$ to $\vertexalt{w}$, but
  by the definition of $\chi$, we infer that $\has(\vertexalt{u})=\has(\vertexalt{w})$.}
\end{proof}}

\paragraph{Tractability results.}
We first show that determining a $k$-BAM can be done efficiently if all preference lengths~$\ell$ are maximum possible, i.e., $\ell=k+1$.
This complements \cref{thm:BAMNPklistlength} where $\ell=k-1$.
The case with $\ell=k$ remains open.%
\begin{restatable}[\appsymb]{proposition}{propBAMpreflen}\label{prop:BAMpreflen}
    If all preference orders in a preference profile are of length $k + 1$, then \bamex{} can be solved in time $|\ppp|^{O(1)}$, i.e., polynomial time.
\end{restatable} 
\appendixproofwithstatement{prop:BAMpreflen}{\propBAMpreflen*}{
\begin{proof}
  Due to \cref{lem:attributes-bound}\eqref{has-bound}, it follows that the \(\cares\) sets of all voters must contain all attributes.
  Due to \cref{lem:rank-dist}, every alternative can only appear at the same rank for all voters (if they appear), as $|\rank w a - \rank v a|$ must be zero for all alternatives $a\in\aaa$ and voter pairs $v,w\in\vvv$. 

  Therefore, by iterating over all preference orders and alternatives in polynomial time, and checking whether an alternative appears in different positions, the problem can be solved in polynomial time.
  If the alternative is found in different positions, the given instance is a NO-instance, by Lemma~\ref{lem:rank-dist}. Otherwise, it is a YES-instance, and the attributes can be assigned trivially based on the rank of an alternative. In this case, it does not matter which of the attributes are assigned, it only matters how many attributes are assigned.
\end{proof}}

The following result shows a dichotomy for the complexity of \bamex{} wrt.\ the number of attributes. \cref{thm:BAMNPklistlength} shows that \bamex\ is NP-hard if there are at least three attributes, while the next theorem shows that for two or less attributes the problem becomes tractable.
\begin{restatable}[\appsymb]{theorem}{thmBAMtwoattpoly}\label{thm:BAMtwoattpoly}
  If $k\leq 2$, \bamex{} is solvable in $O(n)$ time, i.e., linear time.
\end{restatable} 
\appendixproofwithstatement{thm:BAMtwoattpoly}{\thmBAMtwoattpoly*}{
\begin{proof}
  First note that the case $k=1$ is solvable in polynomial time, as preference orders of length one can be ignored, due to only ranking one alternative. Therefore, all preference orders for $k=1$ must have length two, and the problem is polynomial time solvable by \cref{prop:BAMpreflen}, as each preference order has length $k+1=2$. Therefore, it suffices to consider the case of $k=2$. Due to \cref{lem:attributes-bound}\eqref{has-bound}, a profile that contains a preference order of length four or longer is a trivial NO-instance. As preference orders of length one can be ignored, we can assume that any profile only contains preference orders of length two or three.

  We show that such $\bamex$ instances are solvable in linear time, by reducing to \textsc{2-SAT}, which is known to be solvable in linear time~\cite{Krom67}.\probdef{\textsc{2-SAT}}{A set of clauses $\varphi=\{C_1,\ldots,C_{m}\}$ over a set of variables $X=\{x_1,\ldots,x_n\}$, such that for each $C_j\in\varphi$ it holds that $|C_j|=2$.}{Is $\mathcal{C}$ satisfiable?}

  Let (\(\mathcal P\), 2) with $\ppp\coloneqq (\aaa,\vvv,\RR)$ be a \bamex{} instance. We construct a \textsc{2-SAT} instance $I$ such that $\varphi$ is satisfiable if and only if $\ppp$ admits a $2$-BAM. 
	
  From $\ppp$, we construct an instance $I$ of \textsc{2-SAT} as follows:

\begin{compactitem}[--]
\item For every alternative $a\in \aaa$, we add two propositional variable $h_{a,1}$ and $h_{a,2}$ to $X$. These variables will be true, when $a$ \has\ the first respectively second attribute. 
\item For every voter $v\in \vvv$ with a preference order of length three we add the clauses $\phi_{1,v},\ldots,\phi_{6,v}$ to $\varphi$. We assume that the preference order of $v$ has the form $a \succ b \succ c$. The clauses are
  \begin{alignat*}{3}
    \phi_{1,v} \coloneqq~&  h_{a,1} \vee h_{a,1},
    &\quad&                                    
       \phi_{2,v} \coloneqq~&  h_{a,2} \vee h_{a,2},\\
    \phi_{3,v} \coloneqq~ & \neg h_{b,1} \vee \neg h_{b,2}, 
    &&	\phi_{4,v} \coloneqq & h_{b,1} \vee h_{b,2},\\
    \phi_{5,v} \coloneqq~& \neg h_{c,1} \vee \neg h_{c,1}, &&
    \phi_{6,v} \coloneqq~& \neg h_{c,2}\wedge \neg h_{c,2}.
  \end{alignat*}
\item For every voter $v\in \vvv$ with a preference order of length two, we add the clauses $\phi_{7,v},\ldots,\phi_{10,v}$ to $\varphi$. We assume that the preference order of $v$ has the form $d \succ e$. The clauses are
  \begin{alignat*}{3}
    \phi_{7,v} \coloneqq~& h_{d,1} \vee h_{d,2},          
    &\quad&  \phi_{8,v} \coloneqq~& \neg h_{e,1} \vee \neg h_{e,2}, \\
    \phi_{9,v} \coloneqq~& h_{d,1} \vee \neg h_{e,2}, 
    &&\phi_{10,v} \coloneqq~& h_{d,2} \vee \neg h_{e,1}.
  \end{alignat*}
\end{compactitem}
Both $X$ and $\varphi$ can be computed in linear time, as the number of clauses is linear in the number of voters and the number of variables is linear in the number of alternatives. Note, that the number of alternatives can also be considered to be linear in the number of voters, as each alternative that is not ranked by at least one voter can be removed from the instance.

  We now show that $I=(X,\varphi)$ is a YES-instance of \textsc{2-SAT} if and only if $\ppp$ admits a $2$-BAM.

For the `` if'' part, we show that $\varphi$ is satisfiable if $\ppp$ admits a $2$-BAM. W.l.o.g, let $\bamsymb$ be such a $2$-BAM. We show that $\varphi$ is satisfiable. We construct a satisfying assignment of $\varphi$ by setting $h_{a,1}$ (resp. $h_{a,2}$) to true if $a$ \has\ the first attribute (resp. the second attribute) and to false otherwise. We now show that the truth assignment satisfies $\varphi$, by going through all clauses we added.

We first consider the clauses added for voters with preference orders of length three. $\phi_{1,v}$ and $\phi_{2,v}$ need to be satisfied, as the first-ranked alternative must \have\ all attributes, by \cref{lem:attributes-bound}\eqref{has-bound}. Using \cref{lem:attributes-bound}\eqref{has-bound}, it also follows that $\phi_{3,v},\phi_{4,v},\phi_{5,v}$ and $\phi_{6,v}$ must be satisfied, using similar reasoning. Therefore all clauses added for voters with preference order length three are satisfied.

We now consider the clauses added for voters with preference orders of length two. The clause $\phi_{7,v}$ must be satisfied, as per \cref{lem:attributes-bound}\eqref{has-bound} the first-ranked alternative must \have\ at least one attribute. Similarly, per \cref{lem:attributes-bound}\eqref{has-bound} the last-ranked alternative can \have\ at most one attribute, thereby satisfying $\phi_{8,v}$. It must hold that the first-ranked alternative \has\ at least one attribute that the other does not \have\ in order to explain the preference. Therefore if the first-ranked alternative does not \have\ attribute $1$ (resp. attribute $2$), it follows that the other alternative does not \have\ attribute $2$ (resp. attribute $1$). As this implication is identical to the clause $\phi_{9,v}$ (resp. $\phi_{10,v}$), these two clauses must also be satisfied. Therefore, all clauses must be satisfied and we have showed that the variable assignment satisfies $\varphi$.

For the ``only if'' part, we show that $\varphi$ is satisfiable only if $\ppp$ admits a $2$-BAM. Let $\nu\colon X\rightarrow\{1,0\}$ be a truth assignment satisfying $\varphi$. We show that $\ppp$ admits a $2$-BAM. Let $\attrset=\{\alpha,\beta\}$, w.l.o.g.
We construct $\bamsymb \coloneqq (\attrset,\has, \cares)$, such that for each alternative $c \in \aaa$ it holds that
  \begin{alignat*}{3}
    \alpha \in \has{}(c) &\text{, if } \nu(h_{c, 1}) = 1; &\quad&
    \beta \in \has{}(c) &\text{, if } \nu(h_{c, 2}) = 1; \\
    \alpha \not\in \has{}(c) &\text{, if } \nu(h_{c, 1}) = 0;  &\quad&
    \beta \not\in \has{}(c) &\text{, if } \nu(h_{c, 2}) = 0.
  \end{alignat*}
  and for each voter $v \in\vvv$ it holds that $\cares{}(v) \coloneqq \has{}(c^*)$, where $c^*$ is $v$'s most preferred alternative. This assignment of attributes is well-defined and we next show that is constitutes a valid $2$-BAM. 

  We again consider voters with preference orders of length two and three separately.

  Consider a voter with preference order of length three of the form $v\colon a \succ b \succ c$.\@ As $\phi_{1,v}$ and  $\phi_{2,v}$ are satisfied, it follows that $\has{}(a) = \{\alpha, \beta\}$. As $\phi_{3,v}$ and $\phi_{4,v}$ are satisfied, $|\has{}(b)| = 1$. As $\phi_{5,v}$ and $\phi_{6,v}$ are satisfied $\has{}(c)=\emptyset$. By definition, $\cares{}(v) = \{\alpha, \beta\}$, which means that $v$ is explained by $\bamsymb$, since
  \begin{align*}
    \score{v}{a} = 2 > \score{v}{b} = 1 > \score{v}{c} = 0.
  \end{align*}
  Next, consider a voter with preference order of length two of the form $v\colon d \succ e$. Consider the following:
  \begin{compactitem}[--]
  \item As $\phi_{7,v} = h_{d,1} \vee h_{d,2}$ is satisfied, it must be the case that $|\has{}(d)| \ge 1$.
  \item As $\phi_{8,v} = \neg h_{e,1} \vee \neg h_{e,2}$ is satisfied, it must be the case that $|\has{}(e)| \le 1$.
  \item Because of the above and the fact that $\phi_{9,v} = h_{d,1} \vee \neg h_{e,2}$ and $\phi_{10,v} = h_{d,2} \vee \neg h_{e,1}$ are satisfied, it can not be the case that $\has{}(d) = \has{}(e)$, as if both $\textbf{have}$ exactly one attribute, it can not be the same one.
  \end{compactitem}
  As a consequence, given that $\cares{}(v) = \has{}(d)$, $\score{v}{d} > \score{v}{e}$ in all cases.   Thus, every preference order is explained by $\mathcal M$, i.e., profile $\ppp$ admits a $2$-BAM.

This concludes the proof.
\end{proof}}

Next, we consider the number~$m$ of alternatives as parameter.
By \cref{lem:triv-upper-bound} and due to the fact that there are at most $2^m\cdot m!$ many different preference orders,
we immediately obtain a problem kernel wrt.~$m$ (that is, the instance size is bounded by $f(m)$), which yields FPT result for~$m$ by brute-force searching.
Below, we provide an improved approach by only guessing the attribute set for each alternative.

\begin{algorithm}[t]
  \SetAlgoNoEnd
  \SetAlgoLined{}
  \KwIn{A profile~$\ppp = (\aaa, \vvv, \RR)$ with $m=|\aaa|$, $k\in \mathds{N}$}
  \KwOut{\textit{yes} if $\ppp$ admits a $k$-BAM else \textit{no}}
\lIf{$k \ge m(m-1)$}{\Return{yes}}  \label{param-m:sanity}
  \ForEach{$\has \subseteq {2^{[k]}}^m$\label{param-m:attr-c}}{
    \ForEach{$(v,S) \in \vvv\times 2^{[k]}$\label{param-m:attr-v}}{
      \lIf{$(\has,S)$ {\normalfont\text{explains}} $v$}{$\cares{}(v) \leftarrow S$}}
    \lIf{$(\has{}, \cares{})$ is BAM for $\ppp$}{\Return{yes}}
  }
  \Return{no}\label{param-m:bruteforceend}
  \caption{\texttt{BruteForceM}}
  \label{alg:brute-force}
\end{algorithm}

\begin{restatable}[]{theorem}{thmBAMm}\label{thm:BAMm}
    \bamex\ is solvable in time $2^{O(m^3)} \cdot |\mathcal P|^{O(1)}$, which is FPT wrt. $m$. 
\end{restatable} 
\begin{proof}
  The idea is to branch into all possible combinations of attribute subsets for the alternatives
  and for each branch check whether every voter can be explained by choosing an appropriate subset of attributes that he should \care\ about.
  A pseudo-code can be found in \cref{alg:brute-force}.
   
  If $k \ge m(m-1)$, the algorithm is correct due to \cref{lem:triv-upper-bound}. %
  Otherwise, since line~\ref{param-m:attr-c} brute-force searches for all combinations~$\has$ of the attribute subsets for the attributes, we will not miss a correct \has\ function if the instance is a yes-instance.
  In Lines~\ref{param-m:attr-v}--\ref{param-m:bruteforceend} we inspect for each voter whether there exists a subset of attribute that together with the branched $\has$ function can explain the voter's preference order.
  This is correct by the nature of brute forcing and by the fact that the cares functions of different preference orders do not affect each other. 

  It remains to check the running time: Line~\ref{param-m:sanity} runs in constant time.
  After that it is a triply nested for-loop with at most
  $(2^{k})^m \cdot n \cdot 2^{k} \le n \cdot 2^{m^3}$ iterations and a $|\ppp|^{O(1)}$ body.
\end{proof}%

Now, we consider the combined parameter~$(n,k)$.
By \cref{lem:attributes-bound}\eqref{has-bound}, we infer that the sum of the lengths of the preference orders in a profile is at most~$n\cdot (k+1)$.
Since we can ignore alternatives that do not appear in any preference order, we can assume w.l.o.g.\ that
$m\le n\cdot (k+1)$.
Together with \cref{thm:BAMm}, it follows:
\begin{corollary}\label{cor:BAMnk}
\BAMprob\ is solvable in time $2^{O((n\cdot (k+1))^3)} \cdot |\ppp|^{O(1)}$, which is FPT wrt.\ $n+k$.
\end{corollary}

\todoM{Revisit}
As for the single parameter~$n$, our attempts at showing $W[1]$-hardness have been unsuccessful so far.
We weakly conjecture that \BAMprob\ is FPT wrt.\ $n$ and provide a starting point for investigating the problem.

\begin{restatable}[\appsymb]{theorem}{thmBAMPtwovot}\label{thm:BAMPtwovot}
  For $n=2$ voters, one can determine the minimum number~$k$ of attributes of a BAM in $O(m)$ time, i.e., linear time, and compute a corresponding $k$-BAM in $O(m^2)$ time.
\end{restatable} 
\appendixcorrectnessproofwithstatement{%
  \begin{proof}
    Observe that with $2$ voters, there are $3$ different types of attributes, two being cared about by a single voter, and one by both. %
    To determine the minimum~$k$, %
    we first compute a BAM where for each voter~$v$ and each alternative~$c$,
    the~$\score{v}{c}$ is tight, i.e., $\score{v}{c}=|\succ_v| - \rank{v}{c} -1$. 
    We can compute in polynomial time the minimum number of attributes cared about by a single voter, and
    then the minimum number of attributes cared about by both in such a BAM.
    This directly yields the \cares\ function for each voter.
    Afterwards, we compute the \has\ function and show that the computed BAM is an optimal one. 

\mypara{Proof outline.} 
We first define \myemph{attribute types} which play a central role. Afterwards we compute values that keep track of how many attributes of each type each alternative \has. We show correctness of our procedure by showing that these values correspond to a BAM that minimizes the number of attributes.

The correctness proof is split into two parts.
We first show that the computed values represent a BAM with some number~$k$ of attributes that explains the given profile~$\ppp$.
Then, we show that every BAM requires at least $k$~attributes. %

\mypara{The procedure.}
Let $\ppp=(\aaa,\{u, w\}, \RR)$. %

\myunderline{Attribute types and necessary definitions.}  %
We first define the (attribute) types for a given BAM~$\bamsymb=(\attrset, \has, \cares)$. 
The \emph{type} of an attribute $\attr$ is $\type(\attr)\coloneqq\{v\mid \attr\in\cares(v)\}$.
W.l.o.g., we assume that each attribute is cared for by at least one voter in~$\bamsymb$.
For two voters there are hence three types; namely, $\{u\}$, $\{w\}$, and $\{u,w\}$.
Define the set of attributes of each type as $\attrset_u$, $\attrset_w$, and $\attrset_{u,w}$, respectively.
Formally, $\attrset=\attrset_u\uplus\attrset_w\uplus\attrset_{u,w}$, where
$\attrset_u\coloneqq\cares(u)\setminus\cares(w)$, $\attrset_w\coloneqq\cares(w)\setminus\cares(u)$, and  %
$\attrset_{u,w}\coloneqq\cares(u)\cap\cares(w)$.
Similarly, for each alternative~$c$ we partition the attributes it has into $\has_{u,w}$, $\has_u$, and $\has_w$.
Formally, $\has(c)=\has_u(c)\uplus\has_w(c) \uplus \has_{u,w}(c)$, where
\begin{align*}
  &\has_{u}(c) \coloneqq \has{}(c) \cap \attrset_u,\quad \has_{w}(c) \coloneqq \has{}(c) \cap \attrset_w,\\
  & \has_{u, w}(c) \coloneqq \has{}(c) \cap \attrset_{u, w}.
\end{align*}
Finally, for every alternative~$c$ we define some scores: 
$\typescore{u}{c}\coloneqq|\has_u(c)|$, $\typescore{w}{c}\coloneqq|\has_w(c)|$, $\typescore{u,w}{c}\coloneqq|\has_{u,w}(c)|$.
Clearly, $\score{u}{c}=\typescore{u,w}{c}+\typescore{u}{c}$ and $\score{w}{c}=\typescore{u,w}{c}+\typescore{w}{c}$.

\myunderline{Computing the necessary numbers of each attribute type.}
The formal computation steps are provided in \cref{alg:twovot}. We now additionally give an informal description and intuition.
To compute a BAM, the main task is to compute the values~$\typescore{u,w}{c}$, $\typescore{u}{c}$, and $\typescore{w}{c}$ for each alternative~$c\in \aaa$ given above.

Intuitively, these values are lower bounds on the number of the corresponding attribute types that $c$ requires in order to attain its ranks in $u$ and $w$'s preference orders.
Using these values, we compute a (not necessarily unique) BAM, by determining~$\max_{c \in \aaa}\typescore{u}{c}$ attributes of type~$\attrset_u$, $\max_{c \in \aaa}\typescore{w}{c}$ attributes of type $\attrset_w$,
and $\max_{c \in C}\typescore{u,w}{c}$ attributes of type $\attrset_{u,w}$.
Finally, for each alternative~$c$ and each type~$\attrset_T$, we then arbitrarily assign $\typescore{T}{c}$ many attributes of type $\attrset_T$ to alternative $c$.

Recall that for alternative $c$ and voter $v$,  $\score{v}{c}$ corresponds to the number of alternatives that is ranked lower than~$c$ by~$v$, e.g., the last-ranked alternative has score zero.
These values are computed in the for-loop in \cref{line:init}.

The computation then proceeds as follows: %
In the for-loop in \cref{line:bothvot}, we first compute intermediary values, $t_{u,w}$, $t_u$, and $t_w$ for alternatives~$c$ that are ranked by both voters.
These values shall represent the least amount of attributes needed to ensure that $c$ attains its ranks.
In other words, we use as many attributes of type~$\{u, w\}$ as possible, and fill the gap to the corresponding ranks with attributes of type~$\{u\}$ and~$\{w\}$, respectively.
It is intuitive that these ``filler'' attributes are necessary since their role cannot be taken over by attributes of other types.
Since the attribute types are used across the alternatives, in \cref{line:M} we compute the maximum over all filler attributes that only $u$ resp.\ only $v$ \cares\ about, i.e., $M_u\coloneqq\max_{c \in C}t_{u}(c)$ and $M_w\coloneqq\max_{c \in C}t_{w}(c)$.

We remark that %
the $t_{u,w}$-value is not necessarily a lower bound on the number of attributes of type~$\{u,w\}$ since we may reserve more filler attributes of type~$\{u\}$ (resp.\ $\{w\}$) in line~\ref{line:M} and could ``reuse'' them.
Intuitively, if we set the number of attributes of type~$\{u\}$ and $\{w\}$ to $M_u$ and $M_w$, respectively, they are already minimum possible.
Thus it now remains to minimize the number of attributes of type $\{u, w\}$.
The idea in the for-loop in \cref{line:bothvot2}, is thus to determine for each relevant alternative~$c$ the number~$\text{conv}_c$ of attributes that can be reused.
Then we reduce $t_{u, w}(c)$ accordingly to obtain our desired $S_{u, w}(c)$ value, which shall be the minimum number of attributes of type~$\{u,w\}$ needed for~$c$.

Finally, in the for-loops in \cref{line:onevot1,line:onevot2}, the alternatives ranked by only one voter will receive their values. Again, we do so by giving them exactly as many attributes as needed for their position. For alternatives $c\in\aaa$ ranked only by voter $v\in\{u,w\}$, we set $\typescore{v}{c}\coloneqq\min\{M_v,|\succ_v|-\rank{v}{c}-1\}$ and $\typescore{u,w}{c}=\rank{v}{c}-1-\typescore{v}{c}$. Informally, we assign already existing attributes from $\attrset_v$ if possible, and then use attributes from $\attrset_{u,w}$ for the rest and, if necessary, create new attributes in the set $\attrset_{u, w}$. This concludes the informal description.
\begin{algorithm}[t]
  \SetAlgoNoEnd
  \SetAlgoLined{}
  \KwIn{A profile~$\ppp = (\aaa, \{u,w\}, \RR)$}
  \KwOut{Minimum~$k$, and number of attributes~$M_T$ and score values $\typescore{T}{c}$ for all $\emptyset\subset T\subseteq\{u,w\}$ and all~$c\in\aaa$.}
  Initialize all used variables as $0$\;
  \ForEach{$(c,v)\in \aaa\times\vvv$}{\label{line:init}$\lambda_v(c) \coloneqq \begin{cases}
      |\succ_v| - \rank v c -1, & \text{if $v$ ranks $c$} \label{line:rank}\\
      0, & \text{otherwise.} \label{line:0}
    \end{cases}$}
  \ForEach{$c\in C$ that is ranked by both voters\label{line:bothvot}}{%
    \label{line:compute-t}$t_{u, w}(c) \coloneqq \min(\lambda_u(c), \lambda_w(c))$, $t_{u}(c) \coloneqq \lambda_u(c) - t_{u, w}(c)$, $t_{w}(c)\coloneqq \lambda_w(c) - t_{u, w}(c)$}  
  $M_u \coloneqq \max_{c \in \aaa} t_{u}(c)$ and $M_w \coloneqq \max_{c \in \aaa} t_{w}(c)$\label{line:M}
  
  \ForEach{$c\in C$ that is ranked by both voters\label{line:bothvot2}}{$\text{conv}_c \coloneqq \min(M_w - t_{w}(c), M_u - t_{u}(c), t_{u, w}(c))$, \label{line:convert} $\typescore{u,w}{c} \coloneqq t_{u, w}(c) - \text{conv}_c$, $\typescore{u}{c}\coloneqq t_{u}(c) + \text{conv}_{c}$, and $\typescore{w}{c}\coloneqq t_{w}(c) + \text{conv}_{c}$}
  \ForEach{$c\in C$ only ranked by $u$\label{line:onevot1}}{$\typescore{w}{c} \coloneqq 0$, $\typescore{u}{c}\coloneqq \min(\lambda_u(c), M_u)$, $\typescore{u,w}{c}\coloneqq \lambda_u(c) - \typescore{u}{c}$\label{line:u-compute}}
  \ForEach{$c\in C$ only ranked by $w$\label{line:onevot2}}{$\typescore{u}{c} \coloneqq 0$, $\typescore{w}{c}\coloneqq \min(\lambda_w(c), M_w)$, $\typescore{u,w}{c}\coloneqq \lambda_w(c) - \typescore{w}{c}$\label{line:w-compute}}
  $M_{u, w} \coloneqq \max_{c\in \aaa}\typescore{u, w}{c}$; $k \coloneqq M_{u,w}+M_u+M_w$; \label{line:M_{u,w}}

  \Return{$k$, $M_T$ and $S_T(c)$ for all $c\in\aaa$ and $\emptyset\subset T\subseteq\{u,w\}$}
  \caption{\texttt{Value-Computation}}
  \label{alg:twovot}
\end{algorithm}
Note that $k$ is the total number of attributes used by a BAM corresponding to our computed values. Now that we have described the computational steps, it remains to show the correctness of the computed values. In other words, it remains to show that a BAM using a minimum number of attributes can be generated from the computed values.

We defer the complete proof of correctness to \cref{proof:thm:BAMPtwovot} in the supplementary material, but give a brief overview of the correctness proof. It is straightforward to verify that using the values a $k$-BAM can be found. One can simply generate $M_T$ attributes of type $\emptyset\subset T\subseteq\{u,w\}$ and assign $\typescore{T}{c}$ many  arbitrary attributes of type $T$ to alternative $c$. To show that the computed value for $k$ is indeed minimum, we show the existence of two types of alternatives:
\begin{compactitem}[--]
\item An alternative $c$ that must have all attributes in $\attrset_{u, w}$ and all attributes in either $\attrset_{u}$ or $\attrset_w$.
\item An alternative $b$ that must have all attributes in $\attrset_{u}$ (resp. $\attrset_{w}$) and none of the attributes in $\attrset_{w}$ (resp. $\attrset_{u}$).
\end{compactitem} We then show that in order to explain the rankings of these alternatives for the voters, we require at least $k$ attributes. We show this by lower-bounding the score of alternative/voter-pairs and lower-bounding the number of attributes that are in $\has(b)\setminus\has(c)$ and $\has(c)\setminus\has(b)$, respectively.
\end{proof}
}{thm:BAMPtwovot}{\thmBAMPtwovot*}{
\begin{proof}
  Before we continue with the correctness proof, we provide an example below to illustrate how \cref{alg:twovot} works.
  \begin{example}
    Consider a profile with five alternatives and two voters, where the preferences of the voters are as follows:
    \begin{align*}
      u\colon c_1\succ c_2\succ c_3\succ c_4\succ c_5, \quad
      w\colon c_4\succ c_3. 
    \end{align*}
    By the algorithm, in line~\ref{line:init}, we compute that $\lambda_u(c_1)=4$, $\lambda_u(c_2)=3$, $\lambda_u(c_3)=2$, $\lambda_{u}(c_4)=\lambda_w(c_4)=1$, and $\lambda_u(c_5)=\lambda_w(c_3)=0$. 
	In line~\ref{line:bothvot}, we start by considering alternatives that are ranked by both voters.    
	Alternatives $c_3$ and $c_4$ are ranked by both voters and therefore we compute:
    $t_{u,w}(c_3)=0$, $t_{u,w}(c_4)=1$, $t_{u}(c_3)=2$, and $t_w(c_3)=t_u(c_4)=t_w(c_4)=0$.
    Finally, $M_u=2$ and $M_w=0$.
    
    Intuitively, since the scores (i.e., the $\lambda$ values) of alternative~$c_4$ for both voters are equal,
    they can be explained by using only attributes cared for by both voters.
    On the other hand, due to alternative~$c_3$ requiring a higher score of for $u$ than $w$ (i.e., $\lambda_{u}(c_3)-\lambda_{w}(c_3)=2$),
    $c_3$ needs to \have\ at least two attributes that only $u$ \cares\ about.
    
    In line~\ref{line:convert}, we compute that $\text{conv}(c)=0$ and $\typescore{T}{c}=t_T(c)$ for all types~$T$ and all alternatives~$c\in \{c_3,c_4\}$. 
    Intuitively, this means that for~$c_3$ since there are no attributes of type~$\{u,w\}$ (i.e., $t_{u,w}(c_3)=0$), we cannot convert any of these attributes to type~$\{u\}$ or $\{w\}$.
    For $c_4$, since there are no extra attributes of type~$w$ (i.e., $M_w-t_{w}(c_4)=0$) that we can convert to, 
    we cannot convert any attribute of type~$\{u,w\}$ to type~$w$.

    In line~\ref{line:onevot1}, we consider the alternative ranked only by $u$. 
    As for alternative $c_5$, $\lambda_u(c_5)=0$, it follows that $\typescore{T}{c_5}=0$ for all types.
    For alternative $c_2$, $\lambda_u(c_2)=3$. As $M_u=2$, it follows that $\typescore{u}{c_2}=2$ and therefore $\typescore{u,w}{c_2}=1$. For alternative $c_1$, $\lambda_u(c_1)=4$. Once again, it follows that $\typescore{u}{c_1}=2$, as $M_u=2$ and therefore $\typescore{u,w}{c_1}=2$. As there are no alternatives only ranked by $w$, line~\ref{line:onevot2} can be skipped.
    The algorithm returns the previously computed values, $M_u=2$, $M_w=0$, $M_{u,w}=2$, and $k=4$.
    Due to \cref{lem:attributes-bound}(\ref{care-bound}), it already follows that $k=4$ the minimum number of attributes for our example.
    We will show later that it is indeed minimum in general.

    From these computed values, we can compute a BAM.
    We will provide such a BAM in vector notation.
    Since $k=4$, therefore the vectors will have length four.
    Because $M_{u,w}=2$ and $M_u=2$, the first two entries in the vector correspond to the shared attributes in $\attrset_{u, w}$ and the last two entries correspond to the attributes in $\attrset_{u}$. As $M_w=0$, we do not have any attributes in $\attrset_w$.
    \begin{compactitem}[--]
      \item It holds that $\cares(u)=(1,1,1,1)$ and $\cares(w)=(1,1,0,0)$, as $u$ \cares\ about all attributes and $w$ only \cares\ about the shared attributes.
      \item $\has(c_1)=(1,1,1,1)$, as $\typescore{u,w}{c_1}=2$ and $\typescore{u}{c_1}=2$. With similar reasoning we deduce, $\has(c_2)=(0,1,1,1)$, $\has(c_3)=(0,0,1,1)$, $\has(c_4)=(1,0,0,0)$, and $\has(c_5)=(0,0,0,0)$.~\hfill $\diamond$
    \end{compactitem} 
  \end{example}
\mypara{Correctness.}
As $k$ can be determined in linear time, i.e., in $O(m)$ time, it remains to show that
  \begin{compactenum}[(i)]
    \item The computed values correspond to a valid $k$-BAM, which can be computed in linear time. 
    \item The computed values minimize the number of attributes.
\end{compactenum}

\myunderline{Computing $k$-BAM in $O(m^2)$ time.}
We first define a set~$\attrset$ of~$k$ attributes, a \care\ and a \has\ function, and show that it yields a valid~$k$-BAM for~$\ppp$.
\begin{compactitem}[--]
  \item We create a set of $k$ attributes:

  $\attrset\coloneqq$ $\{\attr^u_1,\ldots,\attr^u_{M_u},\attr^w_1,\ldots,\attr^w_{M_w}, \attr^{u,w}_1,\ldots,\attr^{u,w}_{M_{u,w}}\}$. 
\item We set $\cares(u)\coloneqq\{\attr^u_1,\ldots,\attr^u_{M_u}, \attr^{u,w}_1,\ldots,\attr^{u,w}_{M_{u,w}}\}$.
\item We set $\cares(w)\coloneqq\{\attr^w_1,\ldots,\attr^w_{M_w}, \attr^{u,w}_1,\ldots,\attr^{u,w}_{M_{u,w}}\}$.
\item For every alternative $c\in\aaa$, we set $\has(c)\coloneqq$

$\{\attr^u_1,\ldots,\attr^u_{\typescore{u}{c}},\attr^w_1,\ldots,\attr^w_{\typescore{w}{c}}, \attr^{u,w}_1,\ldots,\attr^{u,w}_{\typescore{u,w}{c}}\}$. 
\end{compactitem}
We observe that the generated $\bamsymb$ is tight in terms of the scores.
\begin{observation}\label{obs:scores}
  \begin{compactenum}[(i)]
    \item\label{score-bound} For each alternative~$c\in \aaa$ and type~$\emptyset\subset T\subseteq\{u,w\}$, it holds that $0\le \typescore{T}{c}\le M_{T}$.
    \item\label{score-tight}   
    For each voter~$v\in \{u,w\}$ and each alternative~$c\in \aaa$, it holds that
    $\scoreT{\bamsymb}{v}{c}=\typescore{v}{c}+\typescore{u,w}{c}=\lambda_v(c)=|\succ_u~| - \rank u c -1$.
    \end{compactenum}
\end{observation}
\begin{proof}
  \renewcommand{\qedsymbol}{\hfill~(end of the proof of~\cref{obs:scores}~$\diamond$)}
  \begin{compactenum}[(i)]
    \item That $\typescore{T}{c}\ge 0$ follows directly by checking the three for-loops in lines~\ref{line:bothvot2}, \ref{line:onevot1}, and \ref{line:onevot2}:

    $\text{conv}_c$ is upper-bounded by $t_{u,w}(c)$, and $\typescore{w}{c}$ (resp. $\typescore{u}{c}$) is upper-bounded by $\lambda_w(c)$ (resp. $\lambda_u(c)$), and all terms are non-negative integers.
    Similarly, that $\typescore{T}{c}\le M_{T}$ for $T\neq \{u,w\}$ follows from the definitions in the corresponding case.
    That $\typescore{u,w}{c}\le M_{u,w}$ follows from line~\ref{line:M_{u,w}}.
    \item Consider an arbitrary voter~$v\in \{u,w\}$ and alternative~$c\in \aaa$.
    Clearly, if $c$ is ranked by both voters, then we obtain that $\typescore{v}{c}+\typescore{u,w}{c}\stackrel{\text{line \ref{line:convert}}}{=}t_{v}(c)+t_{u,w}(c) \stackrel{\text{line \ref{line:compute-t}}}{=} \lambda_{v}(c)$.
    If $c$ is ranked by~$v$ only, then we directly obtain that $\typescore{v}{c}+\typescore{u,w}{c} =  \lambda_{v}(c)$ by checking lines~\ref{line:u-compute} and \ref{line:w-compute}, respectively.
    In both cases, we conclude that $\scoreT{\bamsymb}{v}{c}=\typescore{v}{c}+\typescore{u,w}{c}=\lambda_v(c)$.
    Since $\lambda_v(c) = |\succ_u~| - \rank u c -1$ by line~\ref{line:rank}, the statement follows.
  \end{compactenum}
\end{proof}
As $c_1\succ_v c_2$ implies $\rank{u}{c_1}<\rank{u}{c_2}$, \cref{obs:scores} implies that the generated tuple~$(\attrset, \cares,\has)$ is a valid $k$-BAM and explains the preferences of voters $u$ and $w$.
As $k\leq 2m$, it follows directly that this BAM can be computed in $O(m^2)$ time.

\myunderline{Optimality of the computed $k$-BAM.} It remains to show that every BAM that explains the preferences of the voters uses at least $k=M_u+M_w+M_{u,w}$ attributes, assuming that $k>0$ as otherwise nothing needs to be proven.
We do this in three steps:
\begin{compactenum}[(1)]
\item We first show that every alternative $c\in\aaa$ that maximizes $t_u(c)$ (resp. $t_w(c)$) satisfies $\typescore{u}{c}=M_u$ and $\typescore{w}{c}=0$ (resp. $\typescore{w}{c}=M_w$ and $\typescore{u}{c}=0)$.
\item We then show that there exists an alternative~$a\in\aaa$ that 
satisfies $\typescore{u,w}{a}=M_{u,w}$ and one of the two equations ``$\typescore{u}{a}=M_u$ or $\typescore{w}{a}=M_w$''. 
\item Finally, we show that in order to explain the score of the alternatives found in the previous two steps, at least $k$ attributes are needed.
\end{compactenum}

We show the first statement in a claim:
\begin{claim}\label{claim:stat1twopoly}
  \begin{compactenum}[(i)]
    \item\label{claim:Mu>0} If $M_u > 0$, then every alternative~$b_u \coloneqq \argmax_{c \in \aaa} t_{u}(c)$ satisfies $\typescore{u}{b_u}=M_u$, $\lambda_u(c) > 0$, and  $\typescore{w}{b_u}=0$.
    \item \label{claim:Mw>0}If $M_w > 0$, then every alternative $b_w \coloneqq \argmax_{c \in \aaa} t_{w}(c)$ satisfies $\typescore{w}{b_w}=M_w$, $\lambda_w(c) > 0$, and $\typescore{u}{b_w}=0$.
  \end{compactenum}
\end{claim}
\begin{claimproof}{claim:stat1twopoly}
  We only consider the first statement. The second statement works analogously.
  Note that  $M_u = t_{u}(b_u)$ per definition.
  Since $M_u>0$, by line~\ref{line:M}, it follows that $b_u$ is ranked by both voters and so
  \begin{equation}
    \text{conv}_{b_u} = \min(M_w - t_{w}(b_u), M_u - t_{u}(b_u), t_{u, w}(b_u))= 0.\label{conv=0}
  \end{equation}
  By line~\ref{line:convert}, it follows that
  \begin{equation*}
    \typescore{u}{b_u} = t_{u}(b_u) + \text{conv}_{b_u} = t_u(b_u) = M_u. \label{score:bu-Mu}
  \end{equation*}
  Since $M_u>0$, by \cref{obs:scores}\eqref{score-tight}, we immediately obtain that $\lambda_u(c) > 0$. 

  It remains to show that $\typescore{w}{b_u} = 0$.
  Since $t_u(b_u) = M_u>0$, by line~\ref{line:compute-t}, 
  it follows that
  $t_{u}(b_u) = \lambda_u(b_u) - \min(\lambda_u(b_u), \lambda_w(b_u)) > 0$.
  This can hold only if $\lambda_u(b_u) > \lambda_w(b_u)$.
  Then, by the same line,
  we infer that
  \begin{align*}
    \typescore{w}{b_u} & = t_{w}(b_u) + \text{conv}_{b_u}\\
    & \stackrel{\eqref{conv=0}}{=} t_{w}(b_u)\\
    & = \lambda_w(b_u) - \min(\lambda_u(b_u), \lambda_w(b_u)) = 0,
  \end{align*}
  as desired.
    \end{claimproof}
    Next, we show that the second statement holds as each alternative that maximizes $\typescore{u,w}{c}$, must satisfy that no attributes can be converted to attributes in $\attrset_{u}$ or $\attrset_w$:
    \begin{claim}~\label{claim:stat2twopoly}
      Assume that the computed~$k>0$.
      There exists an alternative $a\in\aaa$ for which
    \begin{equation}\label{property:full-house-u}
        \typescore{u,w}{a} = M_{u, w} \text{ and } \typescore{u}{a} = M_u > 0
    \end{equation}
    or 
    \begin{equation}\label{property:full-house-w}
        \typescore{u,w}{a} = M_{u, w} \text{ and } \typescore{w}{a} = M_w > 0
    \end{equation}
    \end{claim}
    \begin{claimproof}{claim:stat2twopoly}
      Since $k>0$, one of $M_{u,w}$, $M_u$, and $M_w$ must be positive.
      If $M_{u,w}=0$ and $M_u > 0$, then let $a\coloneqq \argmax_{c \in \aaa}\typescore{u}{c}$.
      Clearly, by \cref{obs:scores}\eqref{score-bound}, we immediately have $\typescore{u,w}{a} = M_{u, w}$. 

      If $M_{u,w} = 0$ and $M_w>0$, then let $a\coloneqq \argmax_{c \in \aaa}\typescore{w}{c}$.
      By \cref{claim:stat1twopoly}\eqref{claim:Mw>0}, we obtain by our choice of~$a$ that $\typescore{w}{a} = M_w$, as desired.

      Now, we assume that $M_{u,w}>0$. Let $a=\argmax_{c \in \aaa}\typescore{u,w}{c}$.
      By line~\ref{line:M_{u,w}}, it follows that $\typescore{u,w}{a}=M_{u,w}$. We show that $a$ satisfies $\typescore{u}{a}=M_u$ or $\typescore{w}{a}=M_w$. We distinguish between two cases:
      
      \begin{compactenum}
        \item[Case 1: $a$ is ranked by only one voter~$v\in \{u,w\}$.]
        By assumption and by line~\ref{line:onevot1} (resp.\ line~\ref{line:onevot2}), $\typescore{u,w}{a}=M_{u,w}=\lambda_v(a) - \min(\lambda_v(a), M_v) > 0$.
        This implies that $\lambda_v(a)>M_v$.
        By line~\ref{line:onevot1} (resp.\ line~\ref{line:onevot2}), we infer that $\typescore{v}{a}=\min(\lambda_v(a), M_v) = M_v$. %
        
        \item[Case 2: $a$ is ranked by both voters.]
        By assumption and by line~\ref{line:bothvot2}, $0<M_{u,w}=\typescore{u,w}{a}=
        t_{u,w}(a)-\text{conv}_{a}$.
        This implies that  $t_{u,w}(a) > \text{conv}_{a}$.
        By line~\ref{line:convert}, we infer that $t_{u,w}(a) = M_w - t_{w}(c)$ or $t_{u,w}(a) = M_u - t_{u}(c)$.
        Since $\typescore{w}{a} = t_w(a) +  \text{conv}_{a}$ and  $\typescore{u}{a} = t_u(a) +  \text{conv}_{a}$ by line~\ref{line:convert},
        we obtain that $\typescore{w}{a} = M_w$ or $\typescore{u}{a}=M_u$, as desired. 
\end{compactenum}
Since in both cases, $a$ satisfies our required conditions, this concludes the proof.
\end{claimproof}

We can now show that each BAM consists of at least $k$ attributes:
Let $a$ be the alternative described in \cref{claim:stat2twopoly}.
Without loss of generality, assume that it satisfies (\ref{property:full-house-u}), i.e., $\typescore{u,w}{a} = M_{u, w}$ and $\typescore{u}{a} = M_u > 0$.
This also implies that voter~$u$ ranks~$a$.

If $M_w=0$, then $M_u+M_{u,w}=k$ and therefore by \cref{obs:scores}\eqref{score-tight} and \cref{claim:stat2twopoly} that
$k=\typescore{u}{a}+\typescore{u,w}{a}=|\succ_u| - \rank u a -1$; note that $M_u>0$.
Since $|\succ_u| - \rank u a -1=k$ is the smallest possible number to explain $a$ in $u$'s preferences, as argued in \cref{lem:attributes-bound}\eqref{care-bound}, every BAM needs at least $k$ attributes in this case.

Now, we consider the case where $M_w>0$.
Let $b$ be the alternative $b_w$ described in \cref{claim:stat1twopoly}\eqref{claim:Mw>0}. 
We show that in order to explain $a$ and $b$'s ranks in the preferences of $u$ and $w$, we require at least $k$ attributes.

By \cref{obs:scores}\eqref{score-tight} and by our choices of~$a$ and~$b$, we infer that
\begin{align}
  \lambda_u(a) & =\typescore{u}{a}+\typescore{u,w}{a} \nonumber\\
  & = M_u+M_{u,w}\geq\typescore{u}{b}+\typescore{u,w}{b}=\lambda_u(b).\label{eq:pos_u(a)>=pos_u(b)}
\end{align}
To lower-bound the number of attributes, we distinguish between two cases: 
\begin{inparaenum}
\item[Case 1:] $\lambda_w(a)>\lambda_w(b)$, and 
\item[Case 2:] $\lambda_w(a)<\lambda_w(b)$.
\end{inparaenum} 
Note that $\lambda_w(a)=\lambda_w(b)$ is not possible since by our choice of~$b$ we know that $w$ ranks~$b$ and having $\lambda_w(a)=\lambda_w(b)$ would imply that $a$ and $b$ are ranked at the same position in $w$'s preference order.

\begin{compactenum}
  \item[Case 1: $\lambda_w(a) > \lambda_w(b)$.]
  This implies that $w$ must rank $a$, as $\lambda_w(a)$ can not be strictly larger than $\lambda_w(b)$ unless $\lambda_w(a) \ge 1$, which is only the case whenever $w$ ranks~$a$.       

  We first show that
  \begin{align}
    \lambda_w(b) \ge \lambda_u(b).\label{eq:w-b>=u-b}
  \end{align}
  This is clearly the case if $u$ does not rank~$b$ since then $\lambda_u(b)=0$ by line~\ref{line:0}.
  If $u$ ranks~$b$, implying that both voters rank~$b$, then
  by \cref{obs:scores}\eqref{score-tight} and by our choice of~$b$,
  we obtain that
  \begin{align*}
    \lambda_w(b) & = \typescore{w}{b} + \typescore{u,w}{b} \\
    & \stackrel{\text{line~\ref{line:convert}}}{=} t_w(b) + t_{u,w}(b)\\
    & {\ge} t_u(b) + t_{u,w}(b) \stackrel{\text{line~\ref{line:convert}}}{=}   \typescore{u}{b} + \typescore{u,w}{b} =   \lambda_u(b).
  \end{align*}
  Next, we show that
  \begin{align}
    \lambda_u(a) - \lambda_u(b) \ge \lambda_w(a) - \lambda_w(b)\label{eq:gapau>=gapw}
  \end{align}
  This can be shown by considering the difference of the two terms: %
  \begin{align*}
    &~~    \lambda_u(a) - \lambda_u(b) - ( \lambda_w(a) - \lambda_w(b) )\\
    & \ge \typescore{u,w}{a} + \typescore{u}{a} - (\typescore{u,w}{b} + \typescore{u}{b}) -\\
  & ~~~~~  (\typescore{u,w}{a} + \typescore{w}{a}) - (\typescore{u,w}{b} + \typescore{w}{b})\\
    & = \typescore{u}{a} - \typescore{u}{b} + \typescore{w}{b} - \typescore{w}{a}\\
    & = M_u - \typescore{u}{b} + M_w - \typescore{w}{a} \ge 0,
  \end{align*}
  where the first inequality follows from \cref{obs:scores}\eqref{score-tight} and the fact that $\lambda_u(b) \le \typescore{u,w}{b} + \typescore{u}{b}$ (they are equal if $u$ ranks~$b$),
  and the last two inequalities follows from our choices of~$a$ and~$b$.

  Now, we are ready to lower-bound the number attributes needed by all BAMs.
  Informally speaking, every BAM needs to explain the larger gap between~$a$ and~$b$, that is, there must be
  \begin{equation*}
    \max(\lambda_u(a) - \lambda_u(b), \lambda_w(a) - \lambda_w(b)) %
  \end{equation*}
  attributes that $a$ $\has$ and $b$ does not. Otherwise, the gaps between them could not be explained.
      
  In addition to these attributes, by \eqref{eq:w-b>=u-b}, there need to be
  \begin{equation*}
    \max(\lambda_u(b), \lambda_w(b))=\lambda_w(b)
  \end{equation*}
  \emph{other} attributes to explain how $b$ is ranked where it is.
  
  We can therefore lower-bound the number of attributes using the knowledge that the gaps between $a$ and $b$ and $b$'s position for $w$ need to be explained by two disjoint sets of attributes.
  For every $k'$-BAM for $\ppp$, we get that
  \begin{align*}
    &k' \ge\\
    & \max(\lambda_u(a) - \lambda_u(b), \lambda_w(a) - \lambda_w(b)) +  \max(\lambda_u(b),\lambda_w(b))\\
    &\stackrel{\eqref{eq:gapau>=gapw},\eqref{eq:w-b>=u-b}}{=} \lambda_u(a) - \lambda_u(b) +  \lambda_w(b)\\
    & \stackrel{\text{Obs.~\ref{obs:scores}\eqref{score-tight}}}{\ge} \typescore{u,w}{a} + \typescore{u}{a} - (\typescore{u,w}{b} + \typescore{u}{b}) + \\
    &\qquad\quad  (\typescore{u,w}{b} + \typescore{w}{b})\\
    & = M_{u,w} + M_u - S_u(b) + M_w\\
    &= M_{u, w} + M_{u} + M_{w} \\
    &= k,
  \end{align*}
  where the third inquality follows from voter~$u$ ranking~$a$, $\typescore{u,w}{b} + \typescore{u}{b} \ge \lambda_u(b)$ (they are equal if $u$ ranks~$b$), and
  and voter~$w$ ranking~$b$,
  and the fourth and second last equations follow from our choices of~$a$ and~$b$.

  \item[Case 2: $\lambda_w(b) > \lambda_w(a)$.]
  In this case, similar to the previous case, both gaps need to be explained; there need to be $\lambda_u(a) - \lambda_u(b)$
  attributes that $a$ $\has$ but $b$ does not, and 
  $\lambda_w(b) - \lambda_w(a)$ attributes that $b$ $\has$ but $a$ does not. 
  
  In addition, both $b$'s position for $u$ and $a$'s position for $w$ need to be explained.
  These attributes could overlap, therefore we need $\max(\lambda_u(b), \lambda_w(a))$ attributes that are disjoint from the two previously mentioned groups of attributes.

  Putting this together as was done previously, we obtain that for every $k'$-BAM for $\ppp$ 
  \begin{align*}
    &k' \ge\\ &\lambda_u(a) - \lambda_u(b) + \lambda_w(b) - \lambda_w(a) + \max(\lambda_u(b), \lambda_w(a)) \\
    &\ge \lambda_u(a) - \lambda_u(b) + \lambda_w(b) \\
    & \stackrel{\text{Obs.~\ref{obs:scores}\eqref{score-tight}}}{\ge}  (\typescore{u,w}{a} + \typescore{u}{a}) -  (\typescore{u,w}{b} + \typescore{u}{b})\\
    & \quad + (\typescore{u,w}{b} + \typescore{w}{b}) \\
    & = M_{u,w} + M_u - \typescore{u}{b} + M_w\\
    & = M_{u, w} + M_u + M_w \\
    &= k,
  \end{align*}
  where the third inequality follows from voter~$u$ ranking~$a$, $\typescore{u,w}{b} + \typescore{u}{b} \ge \lambda_u(b)$ (they are equal if $u$ ranks~$b$), and
  and voter~$w$ ranking~$b$, and 
  the fourth and fifth inequalities hold by our choices of~$a$ and $b$.
\end{compactenum}
In both cases, we show that $k' \ge k$, as required.

    We have shown that there exists a linear algorithm that computes the least possible amount of attributes for a given preference profile $\ppp$ with $n = 2$. Thus, $\bamex$ is decidable in linear time for $n =2$.
  \end{proof}}

\section{BAM with Cares}\label{sec:BAMwc}
\appendixsection{sec:BAMwc}

In this section we study the restriction on the \BAMprob\ problem, where the \cares-function is given, and the question is whether a \has-function exists that completes a valid $k$-BAM.
Unfortunately, for all single parameters, except for~$n$, this problem remains NP-hard even for constant parameter value.
We complement this by providing FPT-algorithms for all $2$-parameter combinations.

\paragraph{General complexity.}
The next two theorems show that \BAMwcprob\ is NP-complete.
In both cases we reduce from the well-known NP-complete \textsc{3-SAT} problem.
  \toappendix{\probdef{\textsc{3-SAT}}
  {A set $X$ consisting of $n$ variables and a formula $\varphi$ consisting of $m$ clauses of size $3$ over the variables in $X$.}
  {Is there a truth assignment of the variables such that $\varphi$ is true?}}
\begin{restatable}[\appsymb]{theorem}{thmBAMwcNPh}\label{thm:BAMwcNPh}
  \BAMwcprob\ is NP-complete. \BAMwcprob\ remains NP-hard even if $m=3$.
\end{restatable} 
\appendixproofwithstatement{thm:BAMwcNPh}{\thmBAMwcNPh*}{
\begin{proof}

  NP-membership follows from \cref{obs:np-contain}.
  To show NP-hardness, we reduce from the NP-complete \textsc{3-SAT} problem, which results in an instance of \BAMwcprob\ with $m=3$.

  Let $I=(X= \{x_1, \ldots, x_n\},\varphi= \{C_1, \ldots, C_m\})$ be an instance of \textsc{3-SAT}. We create an instance $I'=(\ppp=(\aaa,\vvv,\RR),\cares)$ as follows.

  \mypara{Alternatives.} We create two dummy alternatives $\dummyalt{0}{}$ and $\dummyalt{2}{}$ and a \myemph{choice alternative} $c$. There are $m=3$ alternatives in total.

  \mypara{Voters, their preferences, and their \cares.} 
  \begin{itemize}
    \item[--] For each variable $x\in X$, we add a voter $v_x$ with preference order $v_x\colon \dummyalt{2}{}\succ c\succ \dummyalt{0}{}$ and $\cares(v_x)=\{\attr_{x},\attr_{\neg x}\}$.
	\item[--] For each clause $C\in\varphi$ with $C=\{\ell_1,\ell_2,\ell_3\}$, we add a voter $v_C$ with preference order $v_C\colon c\succ\dummyalt{0}{}$ and $\cares(v_C)=\{\attr_{\ell_1},\attr_{\ell_2},\attr_{\ell_3}\}$.
  \end{itemize}
Note that there are $2n$ attributes in total, as there is two attributes for each variable; one corresponding to the negated variable and one corresponding to the non-negated variable. In other words, $\attrset\coloneqq\{\attr_{x},\attr_{\neg x}\mid x\in X\}$.

  This concludes the construction, which can clearly be done in polynomial time. 
  It remains to show the correctness, i.e., $\varphi$ is satisfiable if and only if the constructed profile~$\ppp$ admits a $k$-BAM.

  For the ``if'' part, let $\bamsymb=(\has, \cares)$ be a $k$-BAM for $\ppp$.
  We aim to show that $\has(c)$ represents a truth assignment that satisfies $\varphi$. First, it must hold that for all variables $x\in X$ $|\{\attr_{x},\attr_{\neg x}\}\cap\has(c)|=1$. This holds, as the voter $v_x$ \cares\ about the attributes $\cares(v_x)=\{\attr_{x},\attr_{\neg x}\}$, and therefore in order to explain the preference order of $v_x$ it must hold that $\score{v_x}{c}=1$, which is equivalent to $|\{\attr_{x},\attr_{\neg x}\}\cap\has(c)|=1$. Therefore we can define a truth assignment $\nu\colon X\to\{0,1\}$ in the following way: \begin{align*}
\nu(x)=\begin{cases}
1\text{, if }\attr_x\in\has(c)\\
0\text{, else}
\end{cases}.
\end{align*}

 We generalize this assignment by saying that $\nu(\neg x)=1-\nu(x)$. In other words, the negated variables truth value is the inverse of the truth value of the unnegated variable.
  
Suppose, for the sake of contradiction that $\nu$ does not satisfy $\varphi$, then there exists a clause $C\in\varphi$, that is not satisfied. If $C=\{\ell_1, \ell_2, \ell_3\}$ is not satisfied, it holds that $\nu(\ell_1)=\nu(\ell_2)=\nu(\ell_3)=0$. From this it follows that $\{\attr_{\ell_1},\attr_{\ell_2},\attr_{\ell_3}\}\cap\has (c)=\emptyset$ and therefore $\score{v_C}{c}=0$. It is not possible that $\bamsymb$ is a valid BAM in this case, as $v_C\colon c\succ\dummyalt{0}{}$ but $\score{v_C}{c}=\score{v_C}{\dummyalt{0}{}}=0$.

  For the ``only if'' part, let $\nu\colon X\to\{0,1\}$ be a truth assignment satisfying $\varphi$. We argue that the following $\has$-function leads to a valid BAM $\bamsymb=(\has,\cares)$:
 Let $\has(\dummyalt{2}{a})=\attrset$, $\has(\dummyalt{0}{})=\emptyset$, and $\has(c)=\{\attr_x\mid x\in X\colon \nu(x)=1\}\cup\{\attr_{\neg x}\mid x\in X\colon \nu(x)=0\}$. For all variables $x\in X$, the preference order of $v_x$ is explained, as $\score{v_x}{\dummyalt{2}{}}=2$, $\score{v_x}{\dummyalt{0}{}}=0$, and $\score{v_x}{c}=1$, as $\nu$ is a valid truth assignment. For each clause $C\in\varphi$, the preference order of voter $v_C$ is explained, as well. This holds, as $\score{v_C}{\dummyalt{0}{}}=0$, and $\score{v_C}{c}\geq 1$. The second inequality must hold, as if $\score{v_C}{c}=0$, it would follow that $\{\attr_{\ell_1},\attr_{\ell_2},\attr_{\ell_3}\}\cap\has (c)=\emptyset$, and therefore $\nu$ would not satisfy clause $C$, a contradiction to our initial assumption.

This concludes the proof.\end{proof}}
Using a similar reduction, we show that even for a constant number of attributes the problem remains NP-hard.
\begin{restatable}[\appsymb]{theorem}{thmBAMwck}\label{thm:BAMwck}
  \BAMwcprob\ remains NP-hard even if $k=6$.
\end{restatable} 
\appendixproofwithstatement{thm:BAMwck}{\thmBAMwck*}{
\begin{proof}
  NP-membership follows from \cref{obs:np-contain}.
  To show NP-hardness, we reduce from the NP-complete \textsc{3-SAT} problem, which results in an instance of \BAMwcprob\ with $k=6$.
  Let $I=(X= \{x_1, \ldots, x_n\},\varphi= \{C_1, \ldots, C_m\})$ be an instance of \textsc{3-SAT}. We create an instance $I'=(\ppp=(\aaa,\vvv,\RR),\cares)$ as follows.

  \mypara{Alternatives.}
\begin{itemize}
\item[--] For each variable $x\in X$ we create an alternative $a_{x}$.
\item[--] For each clause $C\in\varphi$ we create an alternative $a_{C}$.
\item[--] We create four dummy alternatives $\dummyalt{1}{},\ldots,\dummyalt{4}{}$.
\end{itemize}

  \mypara{Voters, their preferences, and their \cares.} 
  \begin{itemize}
    \item[--] For each variable $x\in X$, we add three voters $v_{x,1}$, $v_{x,2}$, and $v_{x,3}$. Voter $v_{x,1}$ has the preference order $v_{x,1}\colon a_x\succ\dummyalt{1}{}$ and $\cares(v_{x,1})=\{\attr_{\textit{Var}}\}$. Voter $v_{x,2}$ has the preference order $v_{x,2}\colon \dummyalt{1}{}\succ a_x\succ \dummyalt{2}{}$ and $\cares(v_{x,2})=\{\attr_T,\attr_F\}$. Voter $v_{x,3}$ has the preference order $v_{x,3}\colon \dummyalt{1}{}\succ \dummyalt{2}{}\succ \dummyalt{3}{}\succ a_x$ and $\cares(v_{x,3})=\{\attr_1,\attr_2,\attr_3\}$.     Intuitively these voters enforce that each alternative corresponding to a variable is assigned exactly one of $\{\attr_T,\attr_F\}$ and $\attr_\textit{Var}$ and none of $\{\attr_1,\attr_2,\attr_3\}$.   

	\item[--] For each clause $C\in\varphi$, we add two voters $v_{C,\textit{Var}}$ and $v_{C,\textit{Num}}$. Voter $v_{C,\textit{Var}}$ has the preference order $v_{C,\textit{Var}}\colon \dummyalt{2}{}\succ a_C$ and $\cares(v_{C,\textit{Var}})=\{\attr_\textit{Var}\}$. Voter $v_{C,\textit{Num}}$ has the preference order $v_{C,\textit{Num}}\colon a_C\succ \dummyalt{4}{}$ and $\cares(v_{C,\textit{Num}})=\{\attr_1,\attr_2,\attr_3\}$. Intuitively, these voters ensure that each alternative corresponding to a clause does not \have\ the attribute $\attr_\textit{Var}$ and contains at least one of the attributes $\{\attr_1,\attr_2,\attr_3\}$.
	\item[--] For each $i\in[3]$ and each clause $C\in\varphi$, we add a voter $v_{C,i}$. Voter $v_{C,i}$ has the preference order $v_{C,i}\colon a_x\succ a_C$, where $x$ is the variable corresponding to the $i$-th literal in $C$. If the $i$-th literal in the clause $C$ is a non-negated variable, $\cares(v_{C,i})=\{\attr_i,\attr_\textit{Var},\attr_T\}$. Else, $\cares(v_{C,i})=\{\attr_i,\attr_\textit{Var},\attr_F\}$.
  \end{itemize}

Note that there are six attributes in total; namely, $\attrset=\{\attr_{1},\attr_2,\attr_3,\attr_\textit{Var},\attr_T,\attr_F\}$. Intuitively, $\attr_T$ and $\attr_F$ is used to assign a truth value to each variable. The numeric attributes $\attr_1$, $\attr_2$, and $\attr_3$,  clarify which of the literals in a clause will be true. The attribute $\attr_\textit{Var}$ is assigned to the alternatives that correspond to a variable in $I$. 

  This concludes the construction, which can clearly be done in polynomial time. 
  It remains to show the correctness, i.e., $\varphi$ is satisfiable if and only if the constructed profile~$\ppp$ admits a $k$-BAM.

  For the ``if'' part, let $\bamsymb=(\has, \cares)$ be a $k$-BAM for $\ppp$. We aim to show that we can construct a truth assignment satisfying $\varphi$ from the \has-function of the alternatives corresponding to variables. We define the truth assignment $\nu\colon X\rightarrow\{0,1\}$ in the following way:
\begin{align*}
\nu(x)\coloneqq\begin{cases}
1\text{, if }\attr_T\in\has(a_x)\\
0\text{, else}
\end{cases}.
\end{align*}  We observe that $|\{\attr_T,\attr_F\}\cap\has(a_x)|=1$ for all $x\in X$, as voter $v_{x,2}$ \cares\ about exactly these attributes, and $\score{v_{x,2}}{a_x}=1$, per \cref{lem:attributes-bound}\eqref{care-bound}. Therefore $\nu$ is well-defined and $\nu(x)=1\Leftrightarrow \attr_T\in\has(a_x)$ and $\nu(x)=0\Leftrightarrow\attr_F\in\has(a_x)$. Further using \cref{lem:attributes-bound}\eqref{care-bound}, we can also infer that $\{\attr_1,\attr_2,\attr_3\}\cap\has(a_x)=\emptyset$. 
Similarly, for each clause $C\in\varphi$, $\attr_\textit{Var}\notin\has(a_C)$, as otherwise the preferences of voter $v_{C,\textit{Var}}$ could not be explained. Finally, $\{\attr_1,\attr_2,\attr_3\}\cap\has(a_C)\neq\emptyset$, for all $C\in\varphi$, due to voter $v_{C,\textit{Num}}$.
  
We show that $\nu$ satisfies every clause $C\in\varphi$. W.l.o.g, let $\attr_1\in\has(a_C)$. Then, $\score{v_{C,1}}{a_C}=1$. Let $a_x$ be the other alternative ranked by $v_{C,1}$. In order to explain $v_{C,1}$'s preference order, it must hold that $\score{v_{C,1}}{a_x}\geq 2$. This can only hold if $\attr_T\in\has(a_x)$, if $x$ appears unnegated in clause $C$ or $\attr_F\in\has(a_x)$, otherwise. Therefore, $\nu$ sets variable $x$ to $1$ if $x$ appears unnegated in clause $C$ and to $0$, otherwise. It follows that $\nu$ satisfies every clause and this direction is shown.

For the ``only if'' part, let $\nu\colon X\to\{0,1\}$ be a truth assignment satisfying $\varphi$. We argue that the following $\has$-function leads to a valid BAM $\bamsymb=(\has,\cares)$:
\begin{itemize}
  \item For every variable $x\in X$, we set $\has(a_x)=\{\attr_\textit{Var},\attr_T\}$ if $\nu(x)=1$ and $\has(x)=\{\attr_\textit{Var},\attr_F\}$, otherwise.
  \item For every clause $C\in\varphi$, it must hold that at least one of the literals in the clause C must be true under $\nu$. Let $i\in[3]$ be an index of a true literal in $C$, e.g., if $C=\{x_2\vee\neg x_4\vee \neg x_7\}$ and the third literal is true, $i=3$. We set $\has(a_C)=\{\attr_i\}$. In the case that multiple literals are true, it suffices to choose either of the true literals' indices.
  \item For the dummies, we set $\has(\dummyalt{1}{})=\{\attr_T,\attr_F,\attr_1,\attr_2,\attr_3\}$, $\has(\dummyalt{2}{})=\{\attr_1,\attr_2,\attr_\textit{Var}\}$, $\has(\dummyalt{3}{})=\{\attr_1\}$, and $\has(\dummyalt{4}{})=\emptyset$.
  \end{itemize}
It can be seen that for every variable $x\in X$:
\begin{itemize}
\item The preference order of $v_{x,1}$ is explained, as $\score{v_{x,1}}{a_x}=1>\score{v_{x,1}}{\dummyalt{1}{}}=0$.
\item The preference order of $v_{x,2}$ is explained, as $\score{v_{x,2}}{\dummyalt{1}{}}=2>\score{v_{x,2}}{a_x}=1>\score{v_{x,2}}{\dummyalt{2}{}}=0$.
\item The preference order of $v_{x,3}$ is explained, as $\score{v_{x,2}}{\dummyalt{1}{}}=3>\score{v_{x,2}}{\dummyalt{2}{}}=2>\score{v_{x,2}}{\dummyalt{3}{}}=1>\score{v_{x,2}}{a_x}=0$.
\end{itemize}
For every clause $C\in\varphi$:
\begin{itemize}
\item The preference order of $v_{C,\textit{Var}}$ is explained, as $\score{v_{C,\textit{Var}}}{\dummyalt{2}{}}=1>\score{v_{C,\textit{Var}}}{a_C}=0$.
\item The preference order of $v_{C,\textit{Num}}$ is explained as $\score{v_{C,\textit{Num}}}{a_C}\geq 1>\score{v_{C,\textit{Num}}}{\dummyalt{4}{}}$.
\item If the $i$-th literal of clause $C$ is not true and $x\in X$ is the variable corresponding to the $i$-th literal, the preference order of $v_{C,i}$ is explained, as $\score{v_{C,i}}{a_x}=1>\score{v_{C,i}}{a_C}=0$.
\item If the $i$-th literal of clause $C$ is true and $x\in X$ is the variable corresponding to the $i$-th literal, the preference order of $v_{C,i}$ is explained, as $\score{v_{C,i}}{a_x}=2>\score{v_{C,i}}{a_C}=1$.
\end{itemize}
 
As $\bamsymb$ explains all the preference orders, it is a valid BAM. This concludes the proof. 
\end{proof}}

\paragraph{Tractability results.}
We now consider the two parameter combinations.
For $(n,m)$, we obtain FPT result via integer linear programming (ILP). %
\begin{restatable}[]{theorem}{thmBAMwcnm}\label{thm:BAMwcnm}
  \BAMwcprob\ can be solved in time \\$(2^n\cdot m)^{O(2^n\cdot m)}\cdot|\ppp|^{O(1)}$, which is FPT wrt.\ $n+m$ .
\end{restatable} 
\begin{proof}
  We solve this problem using an ILP with $O(2^n\cdot m)$ variables and $2^n\cdot m+ n\cdot m$ constraints.
  It is known that ILPs can be solved in FPT time wrt.\ the number of variables~\cite{FrankT87}.
  The intuition behind the ILP is to group the attributes according to types.
  Similarly to \cref{thm:BAMPtwovot}, an attribute's type is defined by the subset of voters that \care\ about it.
  For each attribute type and each alternative we create a variable that stores how many attributes of that type the alternative \has.
  We ensure the preference profile is explained by adding constraints for each voter $v$ and each pair of alternatives that are consecutive in $v$'s preference order.
  We now describe the variables and constraints in the ILP:
  \begin{compactitem}[--]
  \item For every $T\subseteq\vvv$ and every $a\in \aaa$, we add a variable $x_{T,a}$. We compute the number of attributes of type $T$, $m_T=|\{\attr\in\attrset\mid \forall v\in T\colon \attr\in\cares(v)\wedge \forall v\in \vvv\setminus T\colon \attr\notin\cares(v)\}|$ and add the constraint, $x_{T,a}\leq m_T$.
  \item For every voter $v\in \vvv$, we add constraints in the following way. Let $v\colon a_1\succ\cdots\succ a_{m'}$ be the preferences of the voter. We add the following constraint for each $i\in [m'-1]\colon$ $\sum_{v\in T} x_{T,a_{i}}\geq 1+\sum_{v\in T} x_{T,a_{i+1}}$.
  \end{compactitem}

  \noindent Correctness can be checked straightforwardly: The value of each variable corresponds exactly to the number of attributes of that type that alternative $a$ \has.
  If there exists a $k$-BAM with the given~$\cares$, then setting the variables according to the BAM satisfies the constraints.
  On the other hand, if the ILP has a feasible solution~$x_{T,a}$, then we can add $x_{T,a}$ attributes for each type~$T$ to alternative $a$.
  As the number of variables of each type does not exceed the existing attributes of that type and the constraints added for each $i\in [m'-1]$ must be satisfied, this leads to a valid $k$-BAM containing~$\cares$. 
  Since ILP feasibility can be checked in~$O(p^{2.5p+o(p)}\cdot L)$ time, where $L$ is the size of the ILP instance and $p$ the number of variables~\cite{FrankT87}, it follow that our running time is $O((2^n\cdot m)^{2.5\cdot(2^n\cdot m)+o(2^n\cdot m)})\cdot |\ppp|^{O(1)}=(2^n\cdot m)^{O(2^n\cdot m)}\cdot|\ppp|^{O(1)}$.
\end{proof}%
Using \cref{lem:attributes-bound}\eqref{has-bound}, we can then derive the following result from \cref{thm:BAMwcnm}.
\begin{restatable}[\appsymb]{corollary}{thmBAMwcnk}\label{thm:BAMwcnk}
  \BAMwcprob\ is solvable in $(2^n\cdot n\cdot (k+1))^{O(2^n\cdot n\cdot (k+1))}\cdot|\ppp|^{O(1)}$ time, i.e., FPT wrt.\ $n+k$.
\end{restatable} 
\appendixproofwithstatement{thm:BAMwcnk}{\thmBAMwcnk*}{
\begin{proof}
  Similarly to the argumentation for \cref{cor:BAMnk}, we can bound the number of alternatives by $n\cdot (k+1)$. Therefore the result follows directly from \cref{thm:BAMwcnm}.
\end{proof}}
Finally, we can get an FPT algorithm for the parameter $m+k$ by brute-forcing through all possible $\has$ functions. 
\begin{restatable}[\appsymb]{theorem}{thmBAMwcmk}\label{thm:BAMwcmk}
  \BAMwcprob\ can be solved in time $(2^k)^m\cdot |\ppp|^{O(1)}$, which is FPT wrt.\ $m+k$.
\end{restatable} 
\appendixproofwithstatement{thm:BAMwcmk}{\thmBAMwcmk*}{
\begin{proof}
  There exist $(2^k)^m$ many possible $\has$-functions. Per \cref{obs:np-contain}, we can be verify in polynomial time, whether a given $(\cares,\has)$ pair forms a valid BAM. Going through all possible $\has$-functions and checking whether $\bamsymb=(\has,\cares)$ forms a valid BAM leads to an FPT-algorithm with a running time of $(2^k)^m\cdot |\ppp|^{O(1)}$ for a given instance $I$. 
\end{proof}}

\section{BAM with Has}\label{sec:BAMwh}
\appendixsection{sec:BAMwh}
We now study the restriction on the BAM problem, where the $\has$-function is given, and the question is whether a $\cares$-function exists that yields a valid $k$-BAM. We show that the problem is already hard even if there is only one voter, but is FPT wrt.\ the other parameters.
\paragraph{General complexity.}
We show hardness by giving a reduction from the NP-complete \textsc{Restricted Exact 3-set Cover} problem~\cite{RXC3}.

\toappendix{\probdef{\textsc{Restricted Exact 3-set Cover} (RXC3)}
{A set \(X\) of elements with cardinality \(|X|=3 \cdot q\) where \(q \in
    \mathbb{N}\), a collection \(\mathcal{S} = \{S_1, \ldots, S_{3q}\}\),
    where for each subset \(S_i \in \mathcal{S}\), it holds that \(S_i \subset \mathcal{X}\), and \(|S_i| = 3\), and each element is contained in exactly three subsets $S_i$.}
{Is there a collection of subsets \(\mathcal{S'} \subset \mathcal{S}\) such that
    each \(x_i \in X\) is contained in exactly one subset \(S_i' \in
    \mathcal{S'}\)?}
We call a collection of subsets with the above properties a \myemph{restricted exact 3-set cover} for its input. Additionally, note that \(|S'| = q\) follows from the definition of each subset having cardinality exactly
three.}
\begin{restatable}[\appsymb]{theorem}{thmBAMwhNPh}\label{thm:BAMwhNPh}
  \BAMwhprob\ is NP-complete.  \BAMwhprob\ remains NP-hard even if $n=1$.
\end{restatable} 
\appendixproofwithstatement{thm:BAMwhNPh}{\thmBAMwhNPh*}{
\begin{table}[t]
  \caption{Table showing the creation of a \BAMwhprob\ instance from \textsc{RXC3} instance.
    A $1$ in column $c$ and row $\alpha$ indicates that attribute $\alpha \in \has{}(c)$.
    The empty fields depend on the original \textsc{RXC3} instance and are therefore left blank
    in this example.}
  \label{table:cares-existence}
  \begin{center}
    \begin{tabular}{|| c || c | c | c | c | c || c | c | c | c||}
      \hline
      & \(a_1\) & \(a_2\) & \(\dots\) & \(a_{3q-1}\) & \(a_{3q}\) & \(d_2\) & \(d_1\) & \(d_0\) \\
      \hline \hline
      \(\alpha_1\)                         &         &         &           &              &            & 0       & 1       & 0       \\
      \hline
      \(\dots\)                           &         &         &           &              &            & 0       & 1       & 0       \\
      \hline
      \(\alpha_{3q}\)                      &         &         &           &              &            & 0       & 1       & 0       \\
      \hline \hline
      \(\dummyatt{2}{1}\)         & 0       & 1        &  \(\dots\)         & 0            & 0          & 0       & 0       & 0       \\
      \hline
      \(\vdots\)                         & 0       & 0       & \(\ddots\) & 0            & 0          & 0       & 0       & 0       \\
      \hline
      \(\dummyatt{3q-1}{1}\)            & 0       & 0       & \(\dots\)         & 1            & 0          & 0       & 0       & 0       \\
      \hline
      \(\vdots\)                          & 0       & 0       & \(\ddots\)         & 1            & 0          & 0       & 0       & 0       \\
      \hline
      \(\dummyatt{3q-1}{3q-1}\)      & 0       & 0       & \(\dots\)         & 1            & 0          & 0       & 0       & 0       \\
      \hline
      \(\dummyatt{3q}{1}\)            & 0       & 0       & \(\dots\)         & 0            & 1          & 0       & 0       & 0       \\
      \hline
      \(\vdots\)                         & 0       & 0       & \(\ddots\)         & 0            & 1          & 0       & 0       & 0       \\
      \hline
      \(\dummyatt{3q}{3q}\)           & 0       & 0       & \(\dots\)         & 0            & 1          & 0       & 0       & 0       \\
      \hline
      \hline
      \(\dummyatt{3q+1}{1}\)              & 0       & 0       & \(\dots\)         & 0            & 0          & 1       & 0       & 0       \\
      \hline
      \(\vdots\)                         & 0       & 0       & \(\ddots\)         & 0            & 0          & 1       & 0       & 0       \\
      \hline
      \(\dummyatt{3q+1}{3q+2}\)          & 0       & 0       & \(\dots\)         & 0            & 0          & 1       & 0       & 0       \\
      \hline
    \end{tabular}
  \end{center}
\end{table}

\begin{proof}
\newcommand{\exactset}{\ensuremath{\mathcal{E}}}
  NP-membership follows from \cref{obs:np-contain}.
  To show NP-hardness, we reduce from the NP-complete \textsc{RXC3} problem, which results in an instance of \BAMwhprob\ with $n=1$.
  Let \(I=(X=\{x_1,\ldots,x_{3q}\},\mathcal{S})\) be an instance of \textsc{RXC3}. We create an instance $I'=(\ppp=(\aaa,\vvv,\RR),\has)$ as follows.

 \mypara{Alternatives and their attributes.}
\begin{itemize}
\item[--] For each element $x_i\in X$ we create an alternative $a_{i}$.
\item[--] We create three dummy alternatives $\dummyalt{0}{}$, $\dummyalt{1}{}$, and $\dummyalt{2}{}$.
\item[--] For each set $S_j\in \mathcal{S}$, we create an attribute $\alpha_j$.
\item[--] For each element $x_i\in X$ with $i<q$, we create $i-1$ dummy attributes $\dummyatt{i}{1},\ldots,\dummyatt{i}{i-1}$. Note that for $x_1$ no dummy attributes are created.
\item[--] For each element $x_i\in X$ with $i\geq q$, we create $i$ dummy attributes $\dummyatt{i}{1},\ldots,\dummyatt{i}{i}$.
\item[--] We create $3q+2$ dummy attributes $\dummyatt{3q+1}{1},\ldots,\dummyatt{3q+1}{3q+2}$.
\item For each element $x_i\in X$, with $i<q$, we set $\has(a_i)=\{\alpha_j\mid S_j\in\mathcal{S}\colon x_i\in S_j\}\cup\{\dummyatt{i}{1},\ldots,\dummyatt{i}{i-1}\}$.
\item For each element $x_i\in X$, with $i\geq q$, we set $\has(a_i)=\{\alpha_j\mid S_j\in\mathcal{S}\colon x_i\in S_j\}\cup\{\dummyatt{i}{1},\ldots,\dummyatt{i}{i}\}$.
\item For the dummy alternatives, we set $\has(\dummyalt{0}{})=\emptyset$, $\has(\dummyalt{1}{})=\{\alpha_1,\ldots,\alpha_{3q}\}$, and $\has(\dummyalt{2}{})=\{\dummyatt{3q+1}{1},\ldots,\dummyatt{3q+1}{3q+2}\}$.
\end{itemize}

 \mypara{The voter and his preferences.} 
There is one voter $v$ with preference order   \begin{equation*}
    v\colon d_2 \succ a_{3q} \succ \dots \succ a_{q} \succ d_1 \succ a_{q - 1}
    \succ \dots \succ a_1 \succ d_0.
  \end{equation*}

  This concludes the construction, which can clearly be done in polynomial time. See also Table~\ref{table:cares-existence} for an illustration.
  It remains to show the correctness, i.e., \(I=(X=\{x_1,\ldots,x_{3q}\},\mathcal{S})\) is a yes-instance of \textsc{RXC3} if and only if the constructed profile~$\ppp$ admits a \cares-function, such that $(\has,\cares)$ constitutes a BAM.

  For the ``if'' part, let $\bamsymb=(\has, \cares)$ be a $k$-BAM for $\ppp$. We aim to show that the sets corresponding to the attributes $\alpha_j$ in $\has(v)$ form an exact cover.
  We note that the length of the preference order of~$v$ is $|\succ_v|=3q+3$. Therefore, it follows that $\score{v}{\dummyalt{2}{}}=3q+2$, as $|\has(\dummyalt{2}{})|=3q+2$. As the scores in the preference order must be descending it follows that each alternative $b\in\aaa$ has to satisfy $\score{v}{b}=|\succ_v| - \rank v c - 1=3q+2-\rank v c$. It follows that $\score{v}{\dummyalt{1}{}}=q$ and $|\cares(v)\cap\{\alpha_i\mid i\in[3q]\}|=q$. Let $\exactset\coloneqq\{S_j\mid \alpha_j\in\has(v)\}$. We show that $\exactset$ is a restricted exact 3-set cover. Suppose, for the sake of contradiction, that $\exactset$ is not a restricted exact 3-set cover. Then there exists an element $x_i\in X$, such that $x_i\notin\bigcup_{S_j\in\exactset}S_j$. W.l.o.g, assume that $i<q$, then it follows that $\has(a_i)\cap\cares(v)\subseteq\{\dummyatt{i}{1},\ldots,\dummyatt{i}{i-1}\}$. However, then $\score{v}{a_i}\leq i-1<i=q+2-\rank{v}{a_i}$, thereby contradicting our initial observation. As a consequence, $\exactset$ must be a restricted exact 3-set cover. 

  For the ``only if'' part, let $\exactset\subset\mathcal{S}$ be a restricted exact 3-set cover of $X$. For the sake of simplicity, we will refer to the set of dummy attributes as $D\coloneqq\{\dummyatt{i}{1},\ldots,\dummyatt{i}{i-1}\mid i<q\}\cup\{\dummyatt{i}{1},\ldots,\dummyatt{i}{i-1}\mid i\geq q\}\cup\{\dummyatt{3q+1}{1},\ldots,\dummyatt{3q+1}{3q+2}\}$. We argue that $\cares(v)\coloneqq D\cup\{\alpha_j\mid S_j\in\exactset\}$ leads to a valid BAM $\bamsymb=(\has,\cares)$. 
  It is easy to see that $\score{v}{\dummyalt{0}{}}=0$. For every alternative $a_i$, with $i<q$, it holds that $\score{v}{a_i}=i$, as $v$ \cares\ about the $i-1$ dummy attributes that are unique to $a_i$, as well as one attribute $\alpha_j$ that $a_i$ \has, as $\exactset$ is a restricted exact 3-set cover. As $\exactset$ is a restricted exact 3-set cover it must hold that $|\exactset|=q$, and therefore $\score{v}{\dummyalt{1}{}}=q$. For every alternative $a_i$, with $i<q$, it holds that $\score{v}{a_i}=i+1$, as $v$ \cares\ about the $i$ dummy attributes that are unique to $a_i$, as well as caring about one attribute $\alpha_j$ that $a_i$ \has, as $\exactset$ is a restricted exact 3-set cover. Finally, $\score{v}{\dummyalt{2}{}}=q+2$. As these score explain the preference order of $v$, \bamsymb\ is a valid BAM.

This concludes the proof.
\end{proof}}
\paragraph{Tractability results.}
For the parameter~$m$, we can run a separate ILP using $O(2^m)$ many variables to compute the $\cares(v)$ for each voter~$v$ since the \cares-functions do not affect each other.
This immediately yields FPT result for~$m$.
\begin{restatable}[\appsymb]{theorem}{thmBAMwhm}\label{thm:BAMwhm}
  \BAMwhprob\ can be solved in time $(2^m)^{O(2^m)}\cdot|\ppp|^{O(1)}$, which is FPT wrt.\ $m$.
\end{restatable} 
\appendixproofwithstatement{thm:BAMwhm}{\thmBAMwhm*}{
\begin{proof}    
We note that the attributes each voter \cares\ about can be computed separately, as the attributes one voter \cares\ about do not affect another voter. We show that for a given voter, we can determine whether a subset of attributes exist that explains this voter's preference order together with the $\has$ function for the alternatives, in FPT time wrt.\ $m$. By running this algorithm for all voters we can then determine whether there exists a $\cares$ function such that $(\has,\cares)$ is a BAM.
We show this by providing an integer linear program (ILP) with $2^m$ variables and at most $O(2^m)$ constraints. Note that it is known that an ILP is solvable in FPT-time wrt.\ the number of variables~\cite{FrankT87}. The general idea of the ILP is to group the attributes according to types. A type of attribute is defined but the subset of alternatives that \have\ this attribute. As there are $2^m$ possible subsets of alternatives, this number also upper-bounds the number of types we need to consider. For each type $T$ we add a variable, that takes a value between $[0,m_T]$, where $m_T$ is the number of attributes that have this type. To make sure the preference order of the considered voter is explained, we add a constraint for each adjacent pair of alternatives that compares the alternatives' scores.

W.l.o.g, let $v$ be the considered voter and $v\colon a_1\succ a_2\succ\ldots\succ a_{m'}$ be the voter's preferences. We define $\type(\attr)\coloneqq\{a_i\colon \attr\in\has(a_i)\}$. We now describe the formulation of the ILP:
\begin{itemize}
\item For each subset $T\subset\aaa$, we add a variable $v_T$. We also compute the number of attributes that satisfy $\type(\attr)=T$ and save it as an integer $m_T$. We add a constraint that mandates that $v_T\in\{0,\ldots,m_T\}$. 
\item For each $i\in \{1,\ldots,m'-1\}$, we add a constraint $\sum_{a_i\in T} v_T\geq 1+\sum_{a_{i+1}\in T} v_T$.
\end{itemize}
Correctness can be verified easily, as the value of each variable corresponds directly to the number of attributes of the corresponding type that voter $v$ \cares\ about. If there exists a BAM, then setting the variables according to the BAM satisfies the constraints, as the voter can only \care\ about attributes that exist and the scores of higher ranked alternatives are always higher than the score of lower ranked alternatives. On the other hand, if the ILP is a yes-instance we can add arbitrary variables of each type to $\cares(v)$ until there are $v_T$ many of them. As the number of variables of each type does not exceed the existing attributes of that type and the constraints added for each $i\in \{1,\ldots,m'-1\}$ ensure that higher ranked alternatives have a larger score, this leads to a valid BAM. 

ILP feasibility can be checked in time $O(p^{2.5p+o(p)}\cdot L)$, where $L$ is the instance size and $p$ is the number of variables~\cite{FrankT87}. Therefore the running time of this approach is $O((2^m)^{2.5\cdot2^m+o(2^m)}\cdot2^m\cdot|\ppp|^{O(1)})=(2^m)^{O(2^m)}\cdot|\ppp|^{O(1)}$.

This concludes the proof. 
\end{proof}}
\noindent Finally, we can branch over all $2^k$ possibilities of assigning attributes for each of the voters, giving the following result.
\begin{restatable}[\appsymb]{theorem}{thmBAMwhk}\label{thm:BAMwhk}
\BAMwhprob\ can be solved in time $2^k\cdot|\ppp|^{O(1)}$, which is FPT wrt.\ $k$.
\end{restatable} 
\appendixproofwithstatement{thm:BAMwhk}{\thmBAMwhk*}{
\begin{proof}
We note that the attributes each voter \cares\ about can be computed separately, as the attributes one voter \cares\ about do not affect another voter. Therefore, we show that for a given voter, we can determine whether a subset of attributes exist that explains this voter's preferences together with the $\has$ function for the alternatives, in FPT time wrt.\ $k$. By running this algorithm for all voters we can then determine whether there exists a $\cares$ function that such that $(\has,\cares)$ is a $k$-BAM.

For each voter we can branch over all $2^k$ possibilities of assigning attributes to that voter. For each of those assignments we can verify in polynomial time, whether they explain that voter's preferences, per \cref{obs:np-contain}. This approach gives a worst-case running time of $2^k\cdot|\ppp|^{O(1)}$ for an instance $I$, which is FPT wrt.\ $k$. 
\end{proof}}

\section{Conclusion and Open Questions}
On the positive side, we found tractability results if the number~$m$ of alternatives is small or if the number $n$ of voters and the number $k$ of attributes are small.
On the negative side, we had to leave the tractability status with respect to $n$ open and the practically interesting parameterization by $k$ turned out to be intractable.
Since the latter parameterization is related to graph coloring, it seems most promising to consider additional, structural parameters of the instance or solution to obtain more islands of tractability.
For instance, based on a BAM we can define a tri-partite graph consisting of the voters, alternatives, and attributes, representing which voters rank which alternatives, which voters care about which attributes and which alternatives have which attributes.
Perhaps if we restrict the structure of this graph, the problem becomes more tractable?

\clearpage
\section*{Acknowledgments}
The authors are supported by the Vienna Science and Technology Fund (WWTF)~[10.47379/ VRG18012]. We would like to thank the reviewers for their helpful comments.

\bibliography{sample}

\iflong
\clearpage

\setcounter{secnumdepth}{2} 
\renewcommand\thesubsection{\thesection.\arabic{subsection}}

\begin{table}[t!]
  \centering
  \Large \textbf{\appendixtitle}
\end{table}
\bigskip

\begin{appendices}
\appendixtext
\end{appendices}

\fi
\end{document}